\documentclass[12pt,a4paper]{article}
\usepackage[utf8]{inputenc}
\usepackage{amssymb,amsmath,mathrsfs,bm}
\usepackage{comment}
\usepackage{booktabs}
\usepackage{mdframed}
\usepackage{graphicx}
\usepackage{setspace}
\usepackage{a4wide}
\usepackage{tcolorbox}
\usepackage{cite}
\usepackage{hyperref}
\usepackage{enumitem}
\usepackage{bbold}

\usepackage[T1]{fontenc}

\usepackage{tikz}
\usetikzlibrary{calc}
\usetikzlibrary{decorations.pathmorphing}
\usepackage{tikz-feynman}
\usepackage{color}

\makeatletter 
\@addtoreset{equation}{section}
\makeatother

\hypersetup{
  linktocpage,
  colorlinks  = true, 
  urlcolor    = blue, 
  linkcolor   = blue, 
  citecolor   = blue 
}

\def \o{\mathcal{O}}
\newcommand \widebar [1] {\overline{#1}}

\def\yb{\bar{y}}
\def\dbar#1{\widebar{D}_{#1}}
\def\Tr{\text{Tr}}
\def\Zt{\widetilde{Z}}
\def\bh{\widehat\beta}
\def\parm{\partial_\mu}
\def\parmm{\partial^\mu}
\def\Am{A_\mu}
\def\Fmn{F_{\mu\nu}}
\def\Fmnup#1{\ensuremath{F^{(#1)\mu\nu}}}
\def\LLS{L_{\text{LS}}}
\def\A{\mathcal{A}}
\def\C{\mathcal{C}}
\def\F{\mathcal{F}}
\def\L{\mathcal{L}}
\def\Lt{\widetilde{\mathcal{L}}}
\def\N{\mathcal{N}}
\def\P{\mathcal{P}}
\def\Pt{\widetilde{\mathcal{P}}}
\def\Lkin{\mathcal{L}_\text{kin}}
\def\S{\mathcal{S}}
\def\yh{\hat{y}}

\newcommand\abs[1]{\vert#1\vert}
\newcommand{\im}{{\text{i}}}

\usepackage{empheq}
\usepackage{tocloft}
\renewcommand{\arraystretch}{1.2}

\begin{document}

\thispagestyle{empty}

~\\[-2.25cm]

\begin{center}
\vskip 1.7truecm
{\Large\bf
{\LARGE Holographic Correlators Beyond \\\vskip 0.2truecm  Maximal Supersymmetry}
}\\
\vskip 2.25truecm
	{\bf Nikolay Bobev and Hynek Paul \\
	}
		\vskip 0.4truecm
 
	{\it	Instituut voor Theoretische Fysica, KU Leuven, \\
	Celestijnenlaan 200D, B-3001 Leuven, Belgium\\\vskip .2truecm
		\vskip .2truecm   }
	{\it	Leuven Gravity Institute, KU Leuven, \\
	Celestijnenlaan 200D box 2415, 3001 Leuven, Belgium\\\vskip .2truecm
		\vskip .2truecm   }	
		
\vskip .2truecm

\href{nikolay.bobev@kuleuven.be}{{\tt nikolay.bobev@kuleuven.be}} ~~~~ \href{mailto:hynek.paul@kuleuven.be}{{\tt hynek.paul@kuleuven.be}}
\end{center}

\vskip 1.25truecm

\centerline{\bf Abstract}
\vskip .4truecm

\noindent We use the AdS/CFT correspondence to explicitly calculate some of the three-point functions in the planar limit of the 4d $\mathcal{N}=1$ Leigh-Strassler SCFT. This strongly interacting CFT can be obtained as a mass deformation of the 4d $\mathcal{N}=4$ SYM theory and admits a dual description in terms of an AdS$_5$ background of type IIB supergravity. Our analysis is based on the existence of a consistent truncation of the 10d supergravity to a tractable 5d gravitational theory with 10 scalar fields dual to some of the low-lying operators in the spectrum of the LS SCFT. We apply standard holographic techniques to this 10-scalar model to analytically calculate the correlators of interest and thus provide a rare example of explicit three-point functions of scalar non-BPS operators in strongly coupled 4d CFTs. Using superconformal Ward identities we perform several consistency checks of these holographic correlators. As a byproduct of our analysis we discuss some subtleties related to the calculation of extremal correlators in AdS/CFT and the contribution of scalar derivative couplings to the evaluation of Witten diagrams. Our work provides a blueprint for the holographic calculation of other correlators in the LS SCFT and similar top-down holographic models.

\bigskip
\bigskip
	
\begin{center}
\textit{Dedicated to the memory of Ivan T. Todorov.} 

\textit{A pioneer of conformal field theory and a giant of Bulgarian science.}
\end{center}


\newpage
\setcounter{page}{1}\setcounter{footnote}{0}
\setcounter{tocdepth}{2}
\tableofcontents

\section{Introduction}

Explicit calculations of correlation functions of local operators in strongly interacting CFTs in more than two dimensions are notoriously difficult. The AdS/CFT correspondence provides a powerful method to perform such explicit calculations in the planar limit of holographic CFTs. Indeed, since the early days of the holographic correspondence correlation functions of local operators in the maximally supersymmetric SCFTs in 3d, 4d and 6d were computed explicitly by employing the dual supergravity description or world-sheet integrability methods, see \cite{DHoker:2002nbb,Beisert:2010jr} for reviews and further references. These results were later streamlined and generalized in various ways including calculations beyond the leading order in the large-$N$ expansion as well as the development of more efficient Mellin space techniques, see e.g. \cite{Bissi:2022mrs} for a review.

Most of these impressive developments have been focused on top-down holographic theories with maximal supersymmetry, i.e. the SCFTs arising on the worldvolume of coincident M2-, D3-, and M5-branes in flat space. While correlation functions of 4d holographic SCFTs with half-maximal supersymmetry have also been considered in the literature, see for example \cite{Alday:2021odx}, there have been no studies of holographic correlators for top-down SCFTs with less than half-maximal supersymmetry. The goal of this work is to initiate the study of correlations functions in 4d $\mathcal{N}=1$ SCFTs by focusing on a concrete holographic model arising on the worldvolume of D3-branes. 

The holographic calculations of 2pt- and 3pt-functions in maximally supersymmetric SCFTs are special and enjoy various simplifications, as evidenced by the paradigmatic example of the 4d $\mathcal{N}=4$ ${\rm SU}(N)$ SYM theory dual to the AdS$_5\times S^5$ background of type IIB supergravity. The spectrum of KK supergravity modes around this background was computed in \cite{Kim:1985ez} using the coset structure of the $S^5$ manifold and it precisely agrees with the spectrum of BPS operators in the 4d $\mathcal{N}=4$ SYM theory which have integer conformal dimensions that are protected by supersymmetry, i.e. they do not vary as a function of the marginal gauge coupling. Moreover, it was shown in \cite{Lee:1998bxa} that the 3pt-functions of these BPS operators also enjoy a non-renormalization theorem and can be computed in the planar limit either at strong 't Hooft coupling, using Witten diagrams, or at weak coupling using free fields. 

Extending this holographic analysis to 4d $\mathcal{N}=1$ SCFTs presents several calculational and structural difficulties. Even if one has a fully explicit 10d supergravity background, calculating the mass spectrum of supergravity modes is technically complicated and until recently was only possible for internal manifolds admitting a coset description, like the well-known Klebanov-Witten 4d $\mathcal{N}=1$ SCFT \cite{Klebanov:1998hh}. The KK spectrum of its dual AdS$_5\times T^{1,1}$ supergravity solution was computed in \cite{Ceresole:1999zs,Ceresole:1999ht}, see also \cite{Gubser:1998vd}. This calculation derives the spectrum of conformal dimensions of the operators dual to supergravity fields, and thus determines their 2pt-functions. The KK spectrum exhibits a new structural feature not present for the maximally supersymmetric SCFT, namely, some of the KK modes are dual to BPS operators while others belong to long multiplets of the 4d $\mathcal{N}=1$ superconformal algebra and in general have non-rational conformal dimensions. To the best of our knowledge the holographic analysis of 3pt- and higher-point correlation functions in this model has not been performed in the literature.\footnote{The same is true for a large class of holographic 4d $\mathcal{N}=1$ SCFTs arising from D3-branes probing conical singularities of non-compact Calabi-Yau manifolds with a 5d Sasaki-Einstein base. In most of these examples even the spectrum of supergravity KK modes is currently out of reach.} This lack of progress is perhaps due to the complicated analysis needed to expand the type IIB supergravity action to cubic (or higher) order in the fluctuations around the AdS$_5\times T^{1,1}$ background in order to calculate the necessary Witten diagrams.

Our focus in this work will be on a different holographic 4d $\mathcal{N}=1$ SCFT for which we can make progress on these questions. The theory we consider is the Leigh-Strassler SCFT \cite{Leigh:1995ep} which can be thought of as the IR fixed point arising from the 4d $\mathcal{N}=4$ SYM theory after an RG flow triggered by a supersymmetric mass term for one of the three adjoint chiral superfields. As we review in Section~\ref{sec:LS-SCFT}, the IR theory has a (complex) three-dimensional conformal manifold which does not have a free or weakly coupled locus. This in turn implies that the only QFT methods at our disposal for the calculation of correlation functions are restricted to the evaluation of 2pt-functions of operators in short, or BPS, multiplets which are completely fixed by their R-charges and the constraints of conformal invariance. In the planar limit of the LS SCFT holography offers a way out of this impasse. The supergravity dual of the LS SCFT can be explicitly constructed in type IIB supergravity and takes the form of a warped AdS$_5\times \tilde{S}^5$ background with internal fluxes that squash the metric on $S^5$ and break its isometry to ${\rm SU}(2)\times {\rm U}(1)$ \cite{Pilch:2000ej,Pilch:2000fu}. Importantly, this AdS$_5$ background of string theory can be constructed by employing a consistent truncation of type IIB supergravity on $S^5$ to the maximally supersymmetric ${\rm SO}(6)$ gauged supergravity. In the 5d supergravity theory the AdS$_5$ backgrounds dual to the UV $\mathcal{N}=4$ SYM theory and the IR LS SCFT are manifested simply as critical points of the scalar potential \cite{Khavaev:1998fb}. The holographic RG flow that connects these two SCFTs is realized as a supersymmetric domain wall solution that interpolates between the two AdS solutions \cite{Freedman:1999gp}.

The 5d gauged supergravity perspective on this holographic duality provides a way to calculate the correlators of interest. As shown in \cite{Malek:2019eaz} one can use methods from Exceptional Field Theory (ExFT) together with the detailed knowledge of the KK spectrum of the UV AdS vacuum solution to find the KK spectrum of any critical point of the potential in the lower-dimensional gauged supergravity. This method was applied in \cite{Bobev:2020lsk} to compute the full KK spectrum of supergravity excitations of the Pilch-Warner type IIB solution \cite{Pilch:2000ej} and organize them into multiplets of the 4d $\mathcal{N}=1$ superconformal algebra. This amounts to an explicit holographic evaluation of all 2pt-functions of local operators in the LS SCFT dual to supergravity modes.\footnote{See \cite{Muck:2008qnc} for a holographic calculation of some 2pt-functions of low-lying supergravity scalar modes along the holographic RG flows that connects the UV and IR AdS vacua.} As in the Klebanov-Witten theory discussed above, one finds that the KK spectrum contains both short and long multiplets suggesting that this is indeed a structural feature of AdS$_5$/CFT$_4$ holographic pairs with $\mathcal{N}=1$ supersymmetry.

Encouraged by this success, in this work we turn to the holographic calculation of 3pt-functions of scalar operators in the LS SCFT. This is a harder problem since one needs to find a cubic effective action for all supergravity excitations around the Pilch-Warner solution and compute the relevant Witten diagrams. Fortunately, the consistent truncation to 5d gauged supergravity offers a convenient shortcut that allows to solve this problem for a subset of the KK modes dual to the lowest lying operators in the LS SCFT. The 5d maximal gauged supergravity contains 42 scalar fields that have a highly non-trivial and non-linear potential. The fluctuations of these fields around the UV and IR supersymmetric AdS$_5$ vacua described above are dual to the lowest lying scalar operators in the $\mathcal{N}=4$ SYM and $\mathcal{N}=1$ LS SCFTs. The quadratic and cubic terms of these fluctuating scalars in the expansion of the 5d supergravity action encode the conformal dimensions of the dual CFT operators and their 3pt-functions. To simplify this cubic expansion of the 5d theory we focus on a further consistent truncation, derived in \cite{Bobev:2016nua} using the symmetries of the gauged supergravity, which contains 10 scalar fields and is technically more manageable. Importantly, these 10 scalars belong to different superconformal multiplets in the LS SCFT including both BPS and non-BPS operators. 

Employing this 10-scalar model we derive the cubic effective action for the fluctuations around the IR AdS$_5$ solution and then use standard holographic results for Witten diagrams to explicitly compute the 3pt-functions in the dual LS SCFT. These results contain a rare example of analytic calculations of 3pt-functions of non-BPS operators in a strongly coupled 4d CFT. We are not aware of any other method that allows for such a calculation that can be used to test our results. Indeed, currently known supersymmetric localization techniques cannot be applied to the calculation of correlators in 4d $\mathcal{N}=1$ SCFT, while it will be challenging to hone in on the LS SCFT with numerical (super)conformal bootstrap methods due to its low amount of supersymmetry and global symmetry. Despite the absence of any alternative calculational methods to derive our results, we find several consistency checks of the analysis. First, we use the 10-scalar 5d supergravity model to compute the known 3pt-functions of the UV $\mathcal{N}=4$ SYM theory. In addition, we employ two types of 4d $\mathcal{N}=1$ superconformal Ward identities that lead to relations between the 3pt-functions in the LS SCFT and show that our supergravity results exactly reproduce these relations.

In the course of our calculations in the 10-scalar 5d supergravity model we encounter several subtleties. Both in the UV and IR AdS$_5$ solutions we find scalars dual to operators which allow for extremal 3pt-functions, i.e. operators $\mathcal{O}_i$ that have conformal dimensions obeying $\Delta_i+\Delta_j=\Delta_k$. The holographic calculation of such extremal 3pt-functions has been a thorny subject since the early days of AdS/CFT. We discuss the subtlety associated with this calculation, and building on the results of \cite{Aprile:2018efk,Aprile:2020uxk} for the $\mathcal{N}=4$ SYM theory, propose that in general holographic CFTs with such extremal operator arrangements one can define an appropriate notion of ``single-particle operators" for which the extremal 3pt-functions vanish. These single-particle operators in large-$N$ CFTs are obtained as a particular mixture of single- and multi-trace operators determined by imposing orthogonality of their two-point functions with all multi-trace operators in the theory. In addition to the usual cubic couplings arising from the scalar potential of the 5d supergravity, the 10-scalar model yields cubic derivative couplings of the schematic form $\phi(\partial_{\mu} \phi)^2$ which makes the calculation of Witten diagrams somewhat cumbersome. We show that in such setups it is always possible to streamline the analysis and remove the $\phi(\partial_{\mu} \phi)^2$ couplings in favor of pure polynomial $\phi^3$ couplings by adding an appropriate total derivative term to the gravitational Lagrangian. A final subtlety presented by the top-down holographic dual of the LS SCFT is the presence of finite polynomial counter-terms in the supergravity scalars near the AdS$_5$ boundary. These finite boundary terms play an important role in the evaluation of the supergravity on-shell action, see \cite{Bobev:2016nua}, but they also affect the calculation of holographic correlation functions. It turns out that in $\mathcal{N}=4$ SYM and the LS SCFT the finite boundary terms do not affect the 2pt- and 3pt-functions of scalar operators at separate points and only add a contact term $\delta$-function contribution.

We continue in the next section with a summary of some of the properties of the LS SCFT and its conformal manifold. In Section~\ref{sec:10-scalar} we introduce the 5d supergravity 10-scalar consistent truncation and present the non-linear supergravity Lagrangian and its supersymmetric AdS$_5$ solutions. In Section~\ref{sec:3pt_general} we summarize the general procedure for the calculation of 3pt-functions in AdS/CFT, discuss several subtleties related to cubic derivative couplings and extremal correlators and introduce the notion of single-particle operators. Section~\ref{sec:3pt_IR} is devoted to the calculation of the 3pt holographic correlators of the LS SCFT within the 10-scalar supergravity truncation. In Section~\ref{sec:3ptfnctsWard} we present several consistency checks of our holographic results by employing supersymmetric Ward identities and the structure of superconformal blocks in 4d $\mathcal{N}=1$ SCFTs. In Section~\ref{sec:higher-pt} we discuss the holographic calculation of 4pt-functions in the LS SCFT together with a different 5d gauged supergravity consistent truncation, which we use to check our 3pt-function analysis. Some problems for future work and interesting open questions are discussed in Section~\ref{sec:final}. In the two appendices we summarize the free-field and holographic calculation of 3pt- and 4pt-functions in the $\mathcal{N}=4$ SYM theory in the context of our 10-scalar 5d supergravity model.

\section{The LS SCFT}
\label{sec:LS-SCFT}

To define the Leigh-Strassler SCFT \cite{Leigh:1995ep} we start by recalling some properties of the 4d $\mathcal{N}=4$ SYM theory. When presented in $\mathcal{N}=1$ supersymmetry notation the field content of the theory consists of one vector multiplet and three chiral multiplets, $\Phi_{1,2,3}$, that transform in the adjoint representation of the gauge group $G$, which we usually take to be ${\rm SU}(N)$.\footnote{We will often assume that $N\geq 3$ since we are interested in interacting theories and would like to avoid discussing some special features pertaining to the ${\rm SU}(2)$ $\mathcal{N}=4$ SYM theory.} In addition to the canonical K\"ahler potential one needs to add the following superpotential compatible with $\mathcal{N}=4$ supersymmetry
\begin{equation}\label{eq:WN4}
\mathcal{W}_{\mathcal{N}=4} = {\rm Tr}\, \Phi_1[\Phi_2,\Phi_3]\,,
\end{equation}
where the trace is over the adjoint indices as dictated by gauge invariance. This classical Lagrangian defines the $\mathcal{N}=4$ conformal theory in the weakly coupled region of the conformal manifold parametrized by the complexified gauge coupling $\tau = \frac{\theta}{2\pi} +{\rm i} \frac{4\pi}{g_{\rm YM}^2}$. Note that in the $\mathcal{N}=1$ formulation of the $\mathcal{N}=4$ SYM theory only an ${\rm SU}(3)_{\rm F} \times {\rm U}(1)_{\rm R}$ subgroup of the ${\rm SO}(6)$ R-symmetry of the theory is manifest. The three chiral superfields are in the fundamental representation of the ${\rm SU}(3)_{\rm F}$ flavor symmetry which leaves the superpotential~\eqref{eq:WN4} invariant. The ${\rm U}(1)_{\rm R}$ is the canonical R-symmetry present for all $\mathcal{N}=1$ SCFTs and the three chiral superfields have equal R-charges of $\frac{2}{3}$.

\begin{figure}[t]
	\centering
	\includegraphics[trim={2.1cm 3.3cm 6.8cm 3.6cm},clip,width=0.8\textwidth]{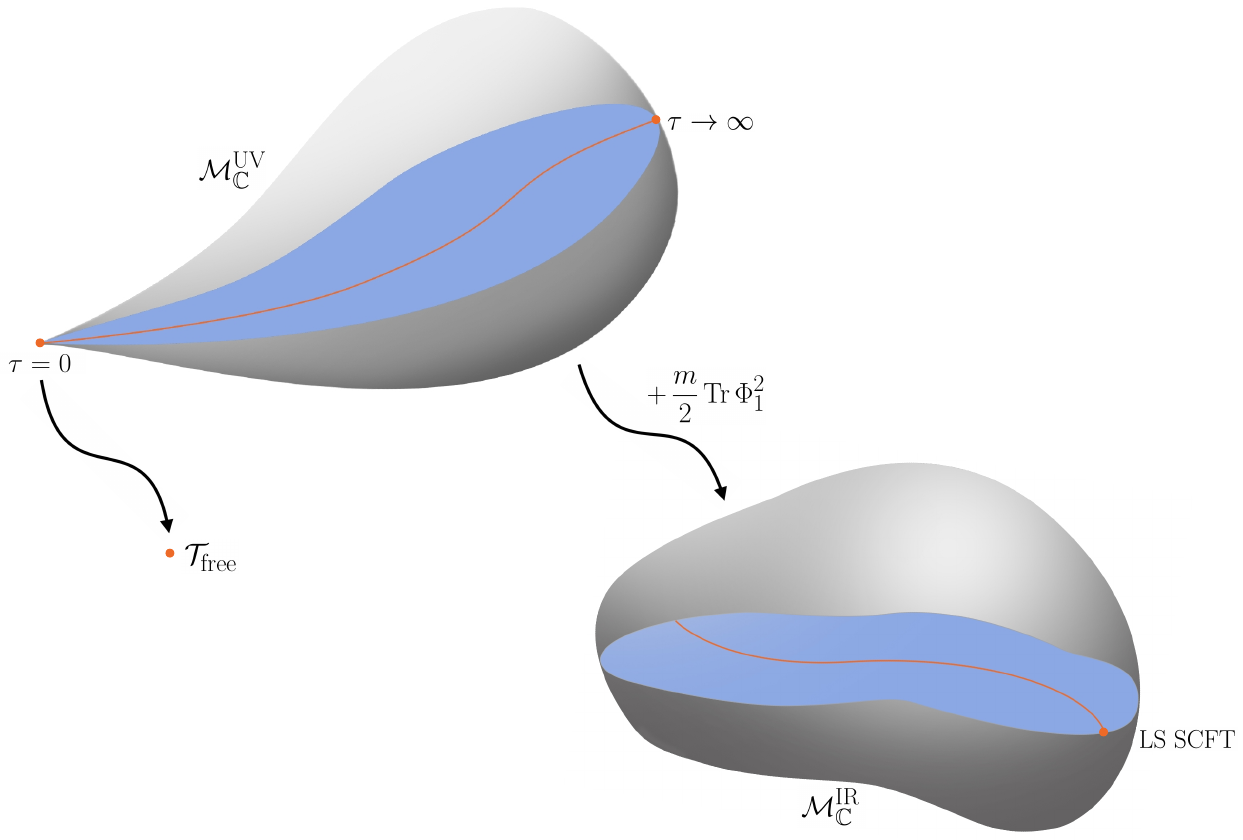}
	\caption{A schematic representation of the UV and IR conformal manifolds and the RG flow between them. The orange lines represent the 1d complex submanifolds of $\mathcal{M}_{\mathbb C}^{\rm UV}$ and $\mathcal{M}_{\mathbb C}^{\rm IR}$ along which the global symmetry is not broken, $\rm SO(6)$ in the UV and ${\rm U(1)_R}\times{\rm SU(2)_F}$ in the IR, respectively. The blue surface is the 2d complex submanifold along which the global symmetry is broken to its maximal Cartan subgroup, i.e. $\rm U(1)^3$ in the UV and ${\rm U(1)_R}\times{\rm U(1)_F}$ in the IR. Finally, on a generic point of the conformal manifolds belonging to the grey region there is no continuous flavor symmetry and only the $\rm U(1)_R$ symmetry of the SCFT is preserved. We hasten to add that this is only a cartoon since not much is known about the global properties of both conformal manifolds.}
	\label{fig:CM}
\end{figure}

As shown in \cite{Leigh:1995ep} there are two additional complex exactly marginal couplings of the $\mathcal{N}=4$ theory that preserve the conformal invariance but break supersymmetry to $\mathcal{N}=1$. A convenient way to present this conformal manifold is given by the superpotential
\begin{equation}\label{eq:WN1UV}
\mathcal{W}_{\mathcal{N}=1}^{\rm UV} = {\rm Tr}\, \Phi_1[\Phi_2,\Phi_3] + h_1 {\rm Tr}\, \Phi_1\{\Phi_2,\Phi_3\} + h_2 {\rm Tr}\left( \Phi_1^3+\Phi_2^3+\Phi_3^3\right) \,.
\end{equation}
The couplings $h_{1,2}$ preserve the ${\rm U}(1)_{\rm R}$ symmetry and break the ${\rm SU}(3)_{\rm F}$ flavor symmetry. For $h_2=0$ the ${\rm U}(1)^2$ subgroup of ${\rm SU}(3)_{\rm F}$ is preserved, while for general values of $h_{1,2}$ there are no continuous flavor symmetries.\footnote{The superpotential \eqref{eq:WN1UV} for general $h_{1,2}$ is invariant under an order $27$ discrete subgroup of ${\rm SU}(3)_{\rm F}$, see Appendix A of \cite{Aharony:2002hx} for further details.} The complex parameters $(\tau,h_1,h_2)$ parametrize the conformal manifold $\mathcal{M}_{\mathbb C}^{\rm  UV}$ of the $\mathcal{N}=4$ SYM theory. While this conformal manifold was originally studied in \cite{Leigh:1995ep} by employing weak coupling $\beta$-function calculations, see also \cite{Aharony:2002hx,Aharony:2002tp}, its existence can also be deduced by other means, for instance by using the superconformal index \cite{Romelsberger:2005eg,Kinney:2005ej} or the method discussed in \cite{Green:2010da,Kol:2010ub,Kol:2002zt}. Not much is known about the global properties of the conformal manifold $\mathcal{M}_{\mathbb C}^{\rm  UV}$. Based on the general result in \cite{Asnin:2009xx} $\mathcal{M}_{\mathbb C}^{\rm  UV}$ should be a K\"ahler manifold, but the corresponding Zamolodchikov metric is not known and it is not clear whether there are any loci (apart from the ones discussed above) with enhanced (super)symmetry or what is the action of S-duality on $\mathcal{M}_{\mathbb C}^{\rm  UV}$. Moreover, little is known about the spectrum of operator dimensions and OPE coefficients at an arbitrary point on the conformal manifold.\footnote{Analytic and numerical conformal bootstrap, supersymmetric localization, holography and integrability methods in the planar limit, have been successfully used to compute some of the correlation functions on the~$\mathcal{N}=4$~submanifold~of~$\mathcal{M}_{\mathbb C}^{\rm  UV}$~with~$h_1=h_2=0$, represented by the orange line in Figure~\ref{fig:CM}.} We present a schematic illustration of $\mathcal{M}_{\mathbb C}^{\rm  UV}$ in Figure~\ref{fig:CM}.

At large $N$, the $\mathcal{N}=4$ SYM theory can be successfully studied using AdS/CFT. The holographic dual of this theory is given by the well-known AdS$_5\times S^5$ background of type IIB string theory. The supergravity dual of the subspace of the conformal manifold with $h_2=0$ was constructed explicitly in \cite{Lunin:2005jy} by utilizing its enhanced ${\rm U}(1)^3$ global symmetry. The supergravity solution holographically dual to a generic point on the conformal manifold $\mathcal{M}_{\mathbb C}^{\rm UV}$ is not known in closed form but some of its properties can be deduced using the results in \cite{Ashmore:2021mao}.

The deformation that leads to the LS SCFT in the IR amounts to adding the following term to the superpotential in \eqref{eq:WN1UV}
\begin{equation}\label{eq:DeltaWLS}
\Delta\mathcal{W} = \frac{m}{2}\,{\rm Tr}\Phi_1^2\,.
\end{equation}
The operator ${\rm Tr}\Phi_1^2$ is relevant and thus the addition of \eqref{eq:DeltaWLS} to the $\mathcal{N}=4$ SYM theory triggers an RG flow. As discussed in \cite{Leigh:1995ep}, the RG flows terminates in an $\mathcal{N}=1$ SCFT in the IR. To understand better the properties of the IR theory it is useful to formally integrate out $\Phi_1$ from the $\mathcal{N}=4$ SYM theory. Using the F-term equations that follow from \eqref{eq:WN1UV} and \eqref{eq:DeltaWLS} one can integrate out $\Phi_1$ and find a quartic superpotential in the IR.\footnote{To find the quartic form of the superpotential after solving the F-term equations for $\Phi_1$ for general values of $h_{1,2}$ one may need to use the chiral ring relations for the UV theory defined by \eqref{eq:WN1UV}. }

Supersymmetry requires that the R-charge of the superpotential is equal to $2$ and thus given its quartic form we can deduce that the two remaining chiral superfields $\Phi_{2,3}$ have R-charge $\frac{1}{2}$ at the IR fixed point. One can form the following seemingly independent gauge invariant quartic combinations of the chiral superfields $\Phi_{2,3}$
\begin{equation}\label{eq:6marginal}
{\rm Tr} \Phi_2^4\,, \qquad {\rm Tr} \Phi_3^4 \,, \qquad  {\rm Tr} \Phi_2^2\Phi_3^2 \,, \qquad  {\rm Tr} \Phi_2\Phi_3\Phi_2\Phi_3 \,, \qquad  {\rm Tr} \Phi_2^3\Phi_3 \,, \qquad  {\rm Tr} \Phi_2\Phi_3^3\,.
\end{equation}
This naively suggests that there are six marginal operators at the IR fixed point. As can be deduced by employing the methods of \cite{Green:2010da,Kol:2010ub,Kol:2002zt} however, not all of these six operators are exactly marginal. It turns out that the complex dimension of the IR conformal manifold, $\mathcal{M}_{\mathbb C}^{\rm IR}$, is three and it can be parametrized formally by the following superpotential 
\begin{equation}\label{eq:WN1LS}
\mathcal{W}_{\mathcal{N}=1}^{\rm IR} = g_1{\rm Tr}\, [\Phi_2,\Phi_3]^2 + g_2 {\rm Tr}\, \{\Phi_2,\Phi_3\}^2 + g_3 {\rm Tr}\left( \Phi_2^4+\Phi_3^4\right) \,,
\end{equation}
in terms of the complex couplings $g_{1,2,3}$. When $g_2=g_3=0$ the superpotential is invariant under $\rm SU(2)_F$ flavor symmetry under which $\Phi_{2,3}$ are in the doublet representation. Indeed, this is one way to find the dimension of the conformal manifold according to the method of \cite{Green:2010da}, i.e. one subtracts the dimension of the flavor group from the number of independent marginal operators in \eqref{eq:6marginal} to find the number of exactly marginal operators.\footnote{One can alternatively use the superconformal index of the LS theory to deduce that there are three complex exactly marginal multiplets in the spectrum, see \cite{Bobev:2020lsk}.} The codimension one locus of $\mathcal{M}_{\mathbb C}^{\rm IR}$ defined by $g_3=0$ and generic values of $g_{1,2}$ has an ${\rm U}(1)_{\rm F}$ flavor symmetry. The generic point on $\mathcal{M}_{\mathbb C}^{\rm IR}$ does not preserve any continuous flavor symmetry.\footnote{The superpotential \eqref{eq:WN1LS} has some discrete flavor symmetries which we will not describe in detail.} Note that this analysis of the conformal manifold of the LS theory is compatible with the calculation of the spectrum of operators in the theory using AdS/CFT. It was shown in \cite{Bobev:2020lsk} that there is one complex marginal operator which is singlet under ${\rm SU}(2)_{\rm F}$ and five complex marginal operators in the $\mathbf{5}$ of ${\rm SU}(2)_{\rm F}$, see the discussion around (2.64) and (2.65) of \cite{Bobev:2020lsk}. This precisely agrees with the list of operators in \eqref{eq:6marginal}.

The global structure of the LS conformal manifold $\mathcal{M}_{\mathbb C}^{\rm IR}$ has not been studied in the literature. In fact, the discussion above based on the superpotential \eqref{eq:WN1LS} is slightly misleading since it suggests that there is a weakly coupled locus on the conformal manifold where one can use a Lagrangian formulation of the LS theory. As we argue below, this is not the case since the conformal and 't Hooft anomalies of the LS CFT are not compatible with the existence of a free locus on $\mathcal{M}_{\mathbb C}^{\rm IR}$. We are therefore led to conjecture that the $\mathcal{M}_{\mathbb C}^{\rm IR}$ conformal manifold is a three-dimensional complex compact space. It will of course be very interesting to learn more about its properties.\footnote{See \cite{Buican:2014sfa} for a discussion on compact conformal manifolds in 4d supersymmetric CFTs.} A schematic illustration of $\mathcal{M}_{\mathbb C}^{\rm  IR}$ is presented in Figure~\ref{fig:CM}.

To compute the $a$ and $c$ conformal anomalies of interest we use the 4d $\mathcal{N}=1$ superconformal Ward identities, see \cite{Anselmi:1997am},
\begin{equation}
a = \frac{9}{32} {\rm Tr} R^3 - \frac{3}{32} {\rm Tr} R\,, \qquad c = \frac{9}{32} {\rm Tr} R^3 - \frac{5}{32} {\rm Tr} R\,,
\end{equation}
where ${\rm Tr} R^3$ and ${\rm Tr} R$ denote the cubic and linear 't Hooft anomaly for the ${\rm U}(1)$ R-symmetry. For $\mathcal{N}=4$ SYM the three chiral superfields have equal R-charge $\frac{2}{3}$. This means that the fermions in these multiplets that contribute to the 't Hooft anomalies have R-charge $-\frac{1}{3}$. The gaugino in the vector multiplet has R-charge $1$. Taking into account the fact that all fermions are in the adjoint representation of the gauge group one then finds the following simple results for the 't Hooft anomalies
\begin{equation}\label{eq:N4tHooft}
{\rm Tr} R^3_{\mathcal{N}=4} = \frac{8}{9}d_{\rm G}\,, \qquad{\rm Tr} R_{\mathcal{N}=4} = 0\,,
\end{equation}
where $d_{\rm G}$ is the dimension of the gauge group. We therefore find the well-known result for the conformal anomalies of the $\mathcal{N}=4$ SYM theory
\begin{equation}\label{eq:N4ac}
a_{\mathcal{N}=4} = c_{\mathcal{N}=4} = \frac{1}{4}d_{\rm G}\,.
\end{equation}

For the LS SCFT we have only 2 chiral superfields $\Phi_{2,3}$ with equal R-charge $\frac{1}{2}$. This means that the fermions in those multiplets have charge $-\frac{1}{2}$. The gaugino in the vector multiplet has the canonical R-charge $1$. This leads to the following 't Hooft anomalies of the R-symmetry of the LS theory
\begin{equation}\label{eq:LStHooft}
{\rm Tr} R^3_{\rm LS} = \frac{3}{4}d_{\rm G}\,, \qquad{\rm Tr} R_{\rm LS} = 0\,. 
\end{equation}
As expected from the general analysis in \cite{Tachikawa:2009tt} the results above amount to the following relation between the UV and IR conformal anomalies
\begin{equation}\label{eq:LSac}
\frac{a_{\rm LS}}{a_{\mathcal{N}=4}} = \frac{c_{\rm LS}}{c_{\mathcal{N}=4}} = \frac{27}{32}\,.
\end{equation}

Since the 't Hooft anomalies above are determined by ${\rm U}(1)$ R-charges they cannot vary continuously over both the UV and the IR conformal manifolds. This in turn implies that the conformal anomalies should also be constant along both $\mathcal{M}_{\mathbb C}^{\rm UV}$ and $\mathcal{M}_{\mathbb C}^{\rm IR}$.\footnote{More generally, the $a$ conformal anomaly cannot change on a conformal manifold due to the $a$-theorem~\cite{Komargodski:2011vj}. If there are non-supersymmetric 4d CFTs with a conformal manifold then the $c$ conformal anomaly could potentially change since it is in general not related to any 't Hooft anomalies and is not necessarily monotonic under RG flows.}  These properties of conformal and 't Hooft anomalies imply that the free point on the UV conformal manifold does not evolve into a point on the IR conformal manifold. To see this more explicitly set $h_1=h_2=0$ in \eqref{eq:WN1UV} and set the gauge coupling $g_{\rm YM}=0$. Adding the mass term \eqref{eq:DeltaWLS} to this free theory and flowing to the IR simply amounts to keeping the R-charges of $\Phi_{2,3}$ to be $\frac{2}{3}$ and removing the superfield $\Phi_1$ from the theory. Thus in the IR, we again have a free theory but with smaller number of fields and 't Hooft anomalies ${\rm Tr} R^3 = \frac{25}{27}d_{\rm G}$ and ${\rm Tr} R = \frac{1}{3}d_{\rm G}$. This is clearly different from the values of the 't Hooft anomalies of the LS theory in \eqref{eq:LStHooft}. We can therefore conclude that the free theory of one vector multiplet and the two adjoint chiral superfields $\Phi_{2,3}$ is not a part of the IR conformal manifold $\mathcal{M}_{\mathbb C}^{\rm IR}$. This is illustrated schematically in Figure~\ref{fig:CM} by the RG flow between the $\tau=0$ point in the UV and the theory denoted by $\mathcal{T}_{\rm free}$ in the IR.

\section{Supergravity truncation: a 10-scalar model}
\label{sec:10-scalar}

The holographic dual of the large-$N$ and strong 't Hooft coupling limit of the $\mathcal{N}=4$ SYM theory is provided by the well-known AdS$_5\times S^5$ background of type IIB supergravity. The gravitational description of the LS mass deformation can be found by exploiting the fact, see \cite{Cvetic:2000nc,Pilch:2000ue,Lee:2014mla,Baguet:2015sma}, that the 10d supergravity theory on $S^5$ admits a consistent truncation to the 5d $\mathcal{N}=8$ ${\rm SO}(6)$ gauged supergravity theory of \cite{Gunaydin:1984qu,Gunaydin:1985cu,Pernici:1985ju}. As shown in \cite{Khavaev:1998fb}, see also \cite{Girardello:1998pd}, the scalar potential of the 5d supergravity theory admits two critical points that correspond to supersymmetric AdS$_5$ vacua representing the holographic dual of the $\mathcal{N}=4$ SYM theory and the LS SCFT.\footnote{There are 32 known AdS$_5$ vacua of the 5d supergravity theory, see \cite{Bobev:2020ttg,Krishnan:2020sfg}. Two of them are supersymmetric and thus perturbatively stable and are the ones we discuss here. The other 30 vacua are non-supersymmetric and suffer from BF instabilities.} Moreover, the full supersymmetric RG flow that connects the $\mathcal{N}=4$ SYM theory and the ${\rm SU}(2)_{\rm F}$ invariant LS SCFT admits a holographic description in terms of a smooth domain wall solution of the 5d gauged supergravity that interpolates between the two supersymmetric vacua \cite{Freedman:1999gp}. The 10d supergravity solution dual to the LS SCFT can be obtained by uplifting the corresponding supersymmetric AdS$_5$ 5d vacuum \cite{Pilch:2000ej}.\footnote{The supersymmetric holographic RG flow solution connecting the UV and IR AdS$_5$ vacua can also be uplifted to type IIB supergravity as shown explicitly in \cite{Pilch:2000fu}.}

To study holographic correlators in the LS SCFT one can employ an explicit 5d consistent truncation of the 10d type IIB supergravity and proceed with the calculation of Witten diagrams. Unfortunately, using the 5d $\mathcal{N}=8$ ${\rm SO}(6)$ gauged supergravity theory for this purpose is not easily tractable since this theory contains all 42 scalar fields dual to scalar operators in the stress-energy tensor multiplet of the $\mathcal{N}=4$ SYM theory. Fortunately there are further consistent truncations of the 5d gauged supergravity that can be constructed by exploiting its symmetries and that have a more manageable number of fields. We will now proceed to study precisely such a truncation constructed in \cite{Bobev:2016nua}, see also \cite{Pilch:2000fu}. Using certain discrete subgroups of the ${\rm SO}(6)\times {\rm SL}(2,\mathbb{R})$ symmetry of the 5d supergravity theory it was shown in \cite{Bobev:2016nua} that the theory can be consistently truncated to a model that contains the 5d metric and 10 real scalar fields. Importantly, both supersymmetric AdS$_5$ vacua of the 5d supergravity theory are solutions of this 10-scalar model. Furthermore, as we discuss in detail below, the 10 scalar fields are dual to operators belonging to four different superconformal multiplets in the LS SCFT and thus provide a rich playground for studying holographic correlators in this theory. We now proceed with the description of the salient features of the 10-scalar model.\footnote{We work with the action derived in \cite{Bobev:2016nua} in Euclidean signature and use a different normalization of the Einstein-Hilbert term. We note that the 10-scalar consistent truncations discussed here and in Section~\ref{sec:higher-pt} are not bosonic sectors of a full 5d $\mathcal{N}=2$ gauged supergravity theory. There are larger, 18-scalar, truncations of the maximal gauged supergravity that comprise fully-fledged 5d $\mathcal{N}=2$ gauged supergravity models coupled to two vectors and four hyper multiplets, see \cite{Bobev:2016nua} and \cite{Bobev:2010de} for further details.}

It is convenient to write the explicit Lagrangian for the 10-scalar model using two real scalars $\beta_{1, 2}$ and four complex fields $z^a$, $a = 1, \ldots, 4$, that parameterize the scalar manifold $(\mathbf{H}^2)^4$. This is a K\"ahler manifold with K\"ahler potential 
\begin{equation}
   \mathcal{K} = -\sum_{a=1}^4 \log(1-z^{a}\bar{z}^{a})\,,
\end{equation}
and K\"ahler metric $\mathcal{K}_{a\bar{b}} \equiv \frac{\partial^2 {\cal K}}{\partial {z^a} \partial {\bar{z}^{ b}}}$.  The relation between the scalars $z^a$ and the 8 scalar fields $\{\varphi,\alpha_{1,2,3},\phi_{1,2,3,4}\}$ that have a direct interpretation in terms of the field theory operators is:
\begin{equation}
  \label{eq:ztophysUV}
  \begin{split}
     z^1 &= \tanh \Big[ \frac{1}{2} \big( \alpha_1 + \alpha_2 + \alpha_3 + \varphi 
       -\im \phi_1 -\im \phi_2 -\im \phi_3 +\im \phi_4 \big) \Big] \,,\\
     z^2 &= \tanh \Big[ \frac{1}{2} \big( \alpha_1 - \alpha_2 + \alpha_3 - \varphi 
       -\im \phi_1 +\im \phi_2 -\im \phi_3 -\im\phi_4 \big) \Big] \,,\\
     z^3 &= \tanh \Big[ \frac{1}{2} \big( \alpha_1 + \alpha_2 - \alpha_3 - \varphi 
       -\im\phi_1 -\im\phi_2 +\im \phi_3 -\im \phi_4 \big) \Big] \,,\\
     z^4 &= \tanh \Big[ \frac{1}{2} \big( \alpha_1 - \alpha_2 - \alpha_3 + \varphi 
       -\im \phi_1 +\im \phi_2 +\im\phi_3 +\im\phi_4 \big) \Big] \,, \\
      \bar z^1 &= \tanh \Big[ \frac{1}{2} \big( \alpha_1 + \alpha_2 + \alpha_3 + \varphi 
       +\im \phi_1 +\im \phi_2 +\im \phi_3 -\im \phi_4 \big) \Big] \,,\\
     \bar z^2 &= \tanh \Big[ \frac{1}{2} \big( \alpha_1 - \alpha_2 + \alpha_3 - \varphi 
       +\im\phi_1 -\im \phi_2 +\im \phi_3 +\im \phi_4 \big) \Big] \,,\\
     \bar z^3 &= \tanh \Big[ \frac{1}{2} \big( \alpha_1 + \alpha_2 - \alpha_3 - \varphi 
       +\im \phi_1 +\im \phi_2 -\im \phi_3 +\im \phi_4 \big) \Big] \,,\\
     \bar z^4 &= \tanh \Big[ \frac{1}{2} \big( \alpha_1 - \alpha_2 - \alpha_3 + \varphi 
       +\im \phi_1 -\im \phi_2 -\im \phi_3 -\im \phi_4 \big) \Big] \,.  
  \end{split}
\end{equation}
In terms of the $z^a$ and $\bar z^{a}$, the gravity-scalar part of the Lagrangian reads
\begin{equation}\label{eq:L_full}
	\L = -R+12(\partial\beta_1)^2+4(\partial\beta_2)^2+ 2\mathcal{K}_{a\bar{b}}\partial_{\mu}z^{a}\partial^{\mu}\bar{z}^{b}+\P\,,
 \end{equation}
where  ${\cal P}$ is the scalar potential.   The scalar potential can be derived from a ``holomorphic superpotential'', i.e. a holomorphic function of the $z_a$, given by 
\begin{equation}\label{eq:suppotdef}
\begin{split}
\mathcal{W} \equiv ~& \frac{1}{L}e^{2\beta_1+2\beta_2}\left(1+z_1z_2+z_1z_3+z_1z_4+z_2z_3+z_2z_4+z_3z_4+z_1z_2z_3z_4\right)\\
&+ \frac{1}{L}e^{2\beta_1-2\beta_2}\left(1-z_1z_2+z_1z_3-z_1z_4-z_2z_3+z_2z_4-z_3z_4+z_1z_2z_3z_4\right)\\
& +\frac{1}{L}e^{-4\beta_1}\left(1+z_1z_2-z_1z_3-z_1z_4-z_2z_3-z_2z_4+z_3z_4+z_1z_2z_3z_4\right),
\end{split}
\end{equation}
through the formula
\begin{equation}\label{eq:10scPot}
   \mathcal{P} = \frac{1}{2}e^{\mathcal{K}}\left[\frac{1}{6}\partial_{\beta_1}\mathcal{W}\partial_{\beta_1}\overline{\mathcal{W}}
     +\frac{1}{2}\partial_{\beta_2}\mathcal{W}\partial_{\beta_2}\overline{\mathcal{W}}+\mathcal{K}^{\bar{b} a}    
      \nabla_{a}\mathcal{W}\nabla_{\bar{b}}\overline{\mathcal{W}} -\frac{8}{3}\mathcal{W}\overline{\mathcal{W}}\right],
 \end{equation}
where $\mathcal{K}^{\bar{b}a}$ is the inverse of ${\cal K}_{a \bar b}$.  The K\"ahler covariant derivative of a function $\mathcal{F}$ is defined as $\nabla_{a} \mathcal{F}\equiv \partial_{a}\mathcal{F} + \mathcal{F}\partial_{a}\mathcal{K}$. The parameter $L$ is related to the gauge coupling of the gauged supergravity and determines the length scale of the AdS$_5$ vacua in the theory. By uplifting these AdS$_5$ solutions to type IIB supergravity and after Dirac quantization of the 10d 5-form flux the 5d gravitational parameters can be related to the number $N$ of D3-branes that source the background. In the Euclidean 5d supergravity action of interest here this amounts to the following overall normalization 
\begin{align}\label{eq:S_eff}
	\S_{\text{eff}}=\eta\int d^5x\sqrt{g}\,\L\,,\qquad\eta=\frac{N^2}{8\pi^2L^3}\,,
\end{align}
where the value of $\eta$ is inherited from the 10-dimensional parent theory upon using $\eta=\frac{1}{16\pi G_5}$ and $G_5=G_{10}/\text{Vol}(S^5)$.

There are two critical points of the potential \eqref{eq:10scPot} that correspond to supersymmetryic AdS$_5$ solutions. They correspond to the holographic dual of the $\mathcal{N}=4$ SYM theory  and the LS SCFT, respectively, and we will therefore refer to them as the UV and IR AdS$_5$ vacua. These take the following form:

\paragraph{UV AdS$_5$ vacuum:}
This solution preserves all supersymmetry and the full ${\rm SO}(6)$ gauge symmetry of the 5d supergravity theory and sits at the origin of the scalar manifold. The values of the 10 scalars and the potential at this critical point are 
\begin{align}\label{eq:sol_UV}
	\varphi=\beta_{1,2}=\alpha_{1,2,3}=\phi_{1,2,3,4}=0\,,\quad\text{with }\mathcal{P}=-\frac{12}{L^2}\,,
\end{align}
where $L$ is the AdS radius. The 10d uplift of this solution is the AdS$_5\times S^5$ background of type IIB supergravity.

\paragraph{IR AdS$_5$ vacuum:}
This solution preserves 8 real supercharges, as appropriate for a holographic dual to a 4d $\mathcal{N}=1$ SCFT, and breaks the gauge group of the 5d supergravity theory to ${\rm SU}(2)\times {\rm U}(1)$. The supergravity scalars acquire non-trivial expectation values given by
\begin{align}\label{eq:sol_IR}
	\varphi=\alpha_{1,2,3}=\phi_{2,3,4}=0\,,\quad\phi_1=-\frac{\pi}{6}\,,\quad\beta_1=\frac{1}{3}\beta_2=\frac{\log(2)}{12}\,,\quad\text{with }\mathcal{P}=-\frac{12}{\LLS^2}\,,
\end{align}
where $\LLS$ is the AdS$_5$ radius of the IR vacuum given by $\LLS=\frac{3}{2^{5/3}}L$. The 10d uplift of this critical point is the Pilch-Warner solution of type IIB supergravity which has a metric given by a warped product of AdS$_5$ and a squashed $S^5$ together with non-trivial internal $B_{(2)}$ and $C_{(2)}$ fluxes \cite{Pilch:2000ej}.

Notice that the ratio of conformal anomalies of the UV and IR SCFT in \eqref{eq:LSac} can be recovered holographically by evaluating the on-shell actions of the two AdS$_5$ vacua
\begin{equation}
\frac{a_{\rm LS}}{a_{\mathcal{N}=4}} = \frac{c_{\rm LS}}{c_{\mathcal{N}=4}} \quad \to \quad \frac{L_{\rm LS}^3}{L^3} = \frac{27}{32}\,.
\end{equation}
We now proceed with a discussion of the spectrum of quadratic fluctuations around the two AdS$_5$ solutions described above.

\subsection{UV spectrum}
\label{subsec:spectrum_UV}

We first consider the scalar mass spectrum of the 10-scalar model around the ${\rm SO}(6)$-invariant UV critical point \eqref{eq:sol_UV}. In order to obtain canonically normalized kinetic terms, we first rescale the fields $\beta_{1,2}$ as follows
\begin{align}
	\beta_1\mapsto\frac{1}{2\sqrt{6}}\beta_1\,,\quad\beta_2\mapsto\frac{1}{2\sqrt{2}}\beta_2\,,\quad\Phi_i\mapsto\frac{1}{2}\Phi_i\,,
\end{align}
where $\Phi_i$ stands for all fields except $\beta_{1,2}$: $\Phi_i\in\{\varphi,\alpha_{1,2,3},\phi_{1,2,3,4}\}$. We then expand the Lagrangian \eqref{eq:L_full} for small values of the scalar fields to find
\begin{equation}\label{eq:L_UV_exp}
\begin{split}
	\L_{\text{kin,UV}} &= \frac{1}{2}\bigg[(\partial_\mu\beta_1)^2+(\partial_\mu\beta_2)^2+(\partial_\mu\varphi)^2+\sum_{i=1}^3(\partial_\mu\alpha_i)^2+\sum_{i=1}^4(\partial_\mu\phi_i)^2\bigg]+\mathcal{L}_{\text{kin,UV}}^{(4)}+\ldots,\\
	\P_{\text{UV}} &= \frac{1}{L^2}\bigg[-12-\frac{1}{2}\Big(4\beta_1^2+4\beta_2^2+4\sum_{i=1}^3\alpha_i^2+3\sum_{i=1}^4\phi_i^2\Big)+\P^{(3)}_{\text{UV}}+\P^{(4)}_{\text{UV}}+\ldots\bigg],
\end{split}
\end{equation}
where $\L_{\text{kin,UV}}^{(n)}$ and $\P_{\text{UV}}^{(n)}$ denote terms which are of $n$-th order in the fields.\footnote{Note that there are no cubic terms in the expansion of $\L_{\text{kin,UV}}$, i.e. $\L_{\text{kin,UV}}^{(3)}=0$. This is due to the fact that we are expanding around the origin of the scalar manifold, but not true for general critical points of the potential. As we discuss later, this is an important subtlety when discussing extremal cubic couplings.}
From the quadratic terms in the expansion of the potential \eqref{eq:L_UV_exp} we can read off the following mass spectrum:
\begin{equation}\label{eq:masses_UV}
\begin{alignedat}{2}
	\alpha_i\,,\beta_i:&\quad m^2L^2=-4\qquad  &\leftrightarrow\qquad \Delta&=2\,,\quad(\text{scalar bilinears, part of}~\mathbf{20}')\\
	\phi_i:&\quad m^2L^2=-3 &\leftrightarrow\qquad \Delta&=3\,,\quad(\text{fermion bilinears, part of}~\mathbf{10}\oplus\widebar{\mathbf{10}})\\
	\varphi:&\quad m^2L^2=0 &\leftrightarrow\qquad \Delta&=4\,.\quad(F_{\mu\nu}F^{\mu\nu},\,\mathbf{1})
\end{alignedat}
\end{equation}
On the right hand side of \eqref{eq:masses_UV} we have presented the conformal dimensions of the operators in the dual $\mathcal{N}=4$ SYM theory, together with the schematic form of these operators in terms of the elementary fields of the SYM theory. We have also indicated the representation of the ${\rm SO}(6)$ R-symmetry to which each of these operators belong to. As anticipated, all 10 scalar fields belong to the 42 scalar operators of the stress-energy tensor multiplet of the $\mathcal{N}=4$ SYM theory.

\subsection{IR spectrum}
\label{subsec:spectrum_IR}

We now turn to the second supersymmetric critical point \eqref{eq:sol_IR}, holographically dual to the LS SCFT. The goal is to again perform an expansion in small field values around this critical point in order to read off the mass spectrum. To arrive at an effective Lagrangian with all quadratic terms in canonical form, we proceed in three steps.

First, in order to describe fluctuations around the IR critical point, it is useful to perform a shift in the zero-point of the fields $\phi_1$ and $\beta_{1,2}$ as follows
\begin{align}\label{eq:step1}
	\quad\phi_1\mapsto-\frac{\pi}{6}-\phi_1\,,\quad\beta_1\mapsto\frac{\log(2)}{12}-\beta_1\,,\quad\beta_2\mapsto\frac{\log(2)}{4}-\beta_2\,,\hspace{0.8cm}
\end{align}
such that the solution \eqref{eq:sol_IR} is obtained by setting all fields to zero.

As a second step, a subsequent rescaling of fields is necessary to bring the kinetic terms at quadratic order into canonical form. This is achieved by 
\begin{equation}\label{eq:step2}
	\beta_1\mapsto\frac{1}{2\sqrt{6}}\beta_1\,,\quad\beta_2\mapsto\frac{1}{2\sqrt{2}}\beta_2\,,\quad\Phi_i\mapsto\frac{\sqrt3}{4}\Phi_i\,,
\end{equation}
where $\Phi_i$ again stands for all other fields except $\beta_{1,2}$. Then, upon inspection of the quadratic terms in the expansion of the potential, one finds that the three fields whose zero-point has been shifted in~\eqref{eq:step1} have acquired a non-diagonal mass matrix, while all other mass terms remain diagonal. The third and last step is therefore to resolve this mixing by performing a rotation in field space. The relevant change of basis from $\{\beta_1,\beta_2,\phi_1\}$ to a new set of fields $\{\bh,\rho_1,\rho_2\}$ reads
\begin{align}\label{eq:step3}
\begin{split}
	\begin{pmatrix}\bh\\\rho_1\\\rho_2\end{pmatrix}\equiv M\cdot \begin{pmatrix}\beta_1\\\beta_2\\\phi_1\end{pmatrix},\quad 
	M=\frac{1}{2\sqrt{14}}\begin{pmatrix}
	-\sqrt{42} & \sqrt{14} & 0 \\
	\sqrt{7+\sqrt{7}} & \sqrt{3(7+\sqrt{7})} & 2\sqrt{7-\sqrt{7}} \\
	-\sqrt{7-\sqrt{7}} & -\sqrt{3(7-\sqrt{7})} & 2\sqrt{7+\sqrt{7}}
	\end{pmatrix}.
\end{split}
\end{align}
where $M$ is an orthogonal matrix.

After all these manipulations, the expansion of the 10-scalar model Lagrangian around the IR AdS$_5$ vacuum is given by 
\begin{align}\label{eq:L_IR_exp}
\begin{split}
	\mathcal{L}_\text{kin,IR}&=\mathcal{L}_\text{kin,IR}^{(2)}+\mathcal{L}_\text{kin,IR}^{(3)}+\mathcal{L}_\text{kin,IR}^{(4)}+\ldots\,,\\[3pt]
	\P_{\text{IR}}&=\frac{1}{\LLS^2}\Big[-12+\P_{\text{IR}}^{(2)}+\P_{\text{IR}}^{(3)}+\P_{\text{IR}}^{(4)}+\ldots\Big],
\end{split}
\end{align}
where the kinetic terms at quadratic order read
\begin{align}
	\mathcal{L}_\text{kin,IR}^{(2)}=\frac{1}{2}\bigg[(\partial_\mu\bh)^2+(\partial_\mu\rho_1)^2+(\partial_\mu\rho_2)^2+(\partial_\mu\varphi)^2+\sum_{i=1}^3(\partial_\mu\alpha_i)^2+\sum_{i=2}^4(\partial_\mu\phi_i)^2\bigg],
\end{align}
and, in contrast to the expansion in the UV, the cubic terms no longer vanish, i.e. $\Lkin^{(3)}\neq0$. The explicit form of these terms will be given in Section \ref{sec:3pt_IR}. The quadratic terms in the expansion of the potential $\P$ take the form
\begin{equation}
	\P^{(2)}_{\rm IR}=\frac{1}{2}\Big[(4-2\sqrt7)\rho_1^2+3\alpha_1^2+(4+2\sqrt7)\rho_2^2-4\bh^2-\tfrac{15}{4}(\alpha_2^2+\alpha_3^2+\phi_2^2+\phi_3^2)-3\phi_4^2\Big],
\end{equation}
from which one can read off the mass spectrum in the IR (in units of $\LLS$), see Table~\ref{tab:PW_spectrum} for a summary. Note that these results are in agreement with the mass spectrum of the 5d maximal gauged supergravity studied in the seminal work \cite{Freedman:1999gp}. Let us now summarize some important features of the IR spectrum:
\begin{table}[t]
\renewcommand{\arraystretch}{1.3}
\begin{center}
$\begin{array}{c|c|c|c|c}
    & m^2\LLS^2 & \Delta & \N=1\text{ multiplet} &SU(2)_{{\rm F}~{U(1)_{\rm R}}}\\\hline
    \alpha_2,\alpha_3 & -\frac{15}{4} & \frac{3}{2} & ~L\bar{B}_1[\frac{3}{2};0,0;1]\otimes[1]^{(1)}+\text{c.c.}~ & \mathbf{3}_1\oplus\mathbf{3}_{-1}\\
    \phi_2,\phi_3 & -\frac{15}{4} & \frac{5}{2} & \text{(chiral)} & \mathbf{3}_{-1}\oplus\mathbf{3}_{1}\\\hline
    \phi_4 & -3 & 3 & L\bar{B}_1[3;0,0;2]\otimes[0]^{(2)}+\text{c.c.} & \mathbf{1}_2\oplus\mathbf{1}_{-2}\\
    \varphi & 0 & 4 &\text{(chiral)} & \mathbf{1}_0\oplus\mathbf{1}_0\\\hline
    \bh & -4& 2 & A\bar{A}[2;0,0;0]\otimes[1]^{(0)} & \mathbf{3}_0\\
    & & & \text{(flavour current)} \\\hline
    \rho_1 & ~4-2\sqrt7~ & ~1+\sqrt7~ & L\bar{L}[1+\sqrt{7};0,0;0]\otimes[0]^{(0)} & \mathbf{1}_0 \\
    \alpha_1 & 3 & 2+\sqrt7 &\text{(long)} & \mathbf{1}_2\oplus\mathbf{1}_{-2} \\
    \rho_2 & 4+2\sqrt7 & 3+\sqrt7 & & \mathbf{1}_0
\end{array}$
\end{center}
\caption{Mass spectrum of the 10-scalar model around the PW AdS$_5$ vacuum. The first, second, and third lines correspond to the $\mathcal{D}(\frac{3}{2},0,0;1)$, $\mathcal{D}(3,0,0;2)$, and $\mathcal{D}(2,0,0;1)$ multiplets in Table 6.1 of \cite{Freedman:1999gp}, respectively. The last line corresponds to the $\mathcal{D}(1+\sqrt{7},0,0;0)$ multiplet in Table 6.2 of \cite{Freedman:1999gp}.}
\label{tab:PW_spectrum}
\end{table}
%
\begin{itemize}
\item A notable fact is that the supergravity spectrum contains modes dual to scalar operators belonging to \textit{long} superconformal multiplets in the LS SCFT. Their corresponding conformal dimensions are irrational numbers. This is a novel feature in the supergravity spectrum, not exhibited by top-down AdS/CFT examples with (half-)maximal supersymmetry.\footnote{The presence of long superconformal multiplets in the supergravity spectrum is familiar from other AdS$_5$/CFT$_4$ examples with $\mathcal{N}=1$ supersymmetry, see for example \cite{Ceresole:1999zs}.}
\item Scalar fields in AdS$_5$ with masses in the range $-4<m^2L^2<-3$ allow for alternate quantization. This is the case for the four scalars $\{\alpha_2,\alpha_3,\phi_2,\phi_3\}$, which have mass $m^2\LLS^2=-\frac{15}{4}$. Fortunately, supersymmetry resolves this ambiguity and determines that $\alpha_{2,3}$ are dual to scalar operators with $\Delta=\frac{3}{2}$, while $\phi_{2,3}$ are dual to scalar operators with $\Delta=\frac{5}{2}$. This is also reflected in the structure of the $L\bar{B}_1[\frac{3}{2};0,0;1]$ superconformal multiplet. Notice that using the superconformal primary in this multiplet that has dimensions $\Delta=\frac{3}{2}$ one can build relevant double-trace operators which can be used to construct a non-supersymmetric IR CFT that can be studied in the planar limit using the results in \cite{Giombi:2017mxl,Giombi:2018vtc}.
\item The 10-scalar model allows us to access only a subset of the 42 scalar fields of the 5d $\mathcal{N}=8$ gauged supergravity. This necessarily means that we do not have enough scalars to reconstruct the full structure of the corresponding superconformal multiplets. This in turn leads to some ambiguity in the identification of bulk scalar fields with superconformal primary (or descendant) operators in the LS SCFT. This can happen when the superconformal multiplet is complex or occurs with multiplicity greater than one due to a non-singlet $\rm SU(2)_F$ representation. In such cases, a given bulk field $\Phi_i$ cannot be assigned to a definite operator in a superconformal multiplet. Instead, $\Phi_i$ is generically mapped to a \textit{linear combination} of the multiplet and its complex conjugate, i.e. the chiral and anti-chiral halves. In the 10-scalar model at hand, this ambiguity concerns the fields $\alpha_{1,2,3}$, $\phi_{2,3,4}$ and $\varphi$. As a direct consequence, these fields do not admit a definite assignment of ${\rm U}(1)$ R-charge, except for $\varphi$, which has vanishing R-charge.
\end{itemize}
From the perspective of the LS SCFT it is useful to organize the supergravity spectrum in terms of $\N=1$ superconformal multiplets. This was already done in \cite{Freedman:1999gp} for the lowest KK-modes on $S^5$, i.e. the supergravity fields that belong to the 5d $\mathcal{N}=8$ gauged supergravity.\footnote{Note that in \cite{Freedman:1999gp} they use a different notation for the 4d $\N=1$ superconformal multiplets from the ones we employ here. In the caption of Table~\ref{tab:PW_spectrum} we provide a comparison between the two conventions.} Recently the full spectrum of higher KK-modes around the Pilch-Warner type IIB supergravity solution was computed and organized in superconformal multiplets in \cite{Bobev:2020lsk}. Following the conventions of \cite{Bobev:2020lsk}, we denote the 4d $\N=1$ superconformal multiplets by
\begin{align}\label{eq:XYSCmult}
	X\bar{Y}[\Delta;j_1,j_2;r]\otimes[k]^{(\yh)}\,,
\end{align}
where the letters $X$ and $Y$ denote the type of shortening conditions enjoyed by the multiplet, see \cite{Cordova:2016emh} for more details on the 4d $\mathcal{N}=1$ superconformal representation theory. The other labels in \eqref{eq:XYSCmult} denote the quantum numbers of the superconformal primary operator: $\Delta$ is the conformal dimension, $j_1$ and $j_2$ the Lorentz spin, $r$ the $U(1)_{\rm R}$ charge, and $k$ is the $\rm SU(2)_F$ spin. The superscript $\yh$ does not indicate a symmetry quantum number but was found to be useful in the organization of the KK-spectrum in \cite{Bobev:2020lsk}. The value of $\yh\equiv p+2y$ is determined by the charge $p$ under the ${\rm U}(1)$ subgroup of ${\rm SO}(6)$ broken by the LS mass term \eqref{eq:DeltaWLS}, while $y$ is the charge under the ``bonus'' ${\rm U}(1)_{Y}$ symmetry arising in the large-$N$ limit of the $\N=4$ SYM theory, \cite{Intriligator:1998ig}.

\section{Holographic 3pt-functions: generalities}
\label{sec:3pt_general}

Before going into the explicit computation of the holographic 3pt-functions for the 10-scalar model introduced above, here we review the general procedure for the calculation of holographic 3pt-functions from bulk cubic couplings assuming a generic gravitational effective Lagrangian $\L_{\text{eff}}$. This will allow us to comment on some often overlooked subtleties regarding field redefinitions and extremal couplings. Schematically, the holographic prescription for the calculation of 3p-functions is a map of the form
\begin{center}
\begin{tabular}{c}
	bulk cubic couplings\\
	$c_{ijk}\,\Phi_i\Phi_j\Phi_k+d_{ijk}\,\Phi_i\nabla_{\mu}\Phi_j\nabla^{\mu}\Phi_k\subset\L_{\text{eff}}$
\end{tabular}
$\quad\mapsto\quad$
\begin{tabular}{c}
	CFT 3pt-functions\\
	$\langle\o_{\Phi_i}\o_{\Phi_j}\o_{\Phi_k}\rangle$
\end{tabular}\,.
\end{center}
The 3pt-functions $\langle\o_{\Phi_i}\o_{\Phi_j}\o_{\Phi_k}\rangle$ are completely fixed by conformal invariance in terms of the conformal dimensions of the operators and the 3pt-function coefficients $C_{\Phi_i\Phi_j\Phi_k}$, which are determined holographically in terms of the supergravity couplings $c_{ijk}$ and $d_{ijk}$. As we discuss below, there is a subtlety in how to translate the bulk cubic couplings to the CFT coefficients $C_{\Phi_i\Phi_j\Phi_k}$. It turns out that by adding suitable total derivative terms to the effective action $\L_{\text{eff}}$, all derivative-couplings $d_{ijk}$ can be removed without altering the CFT observables $C_{\Phi_i\Phi_j\Phi_k}$. We also emphasize that the calculation of extremal 3pt-functions is inherently ambiguous and relate this ambiguity to operator mixing in the dual CFT picture.

\subsection{From cubic couplings in the bulk to CFT 3pt-functions}
\label{subsec:bulkcubic}
As a starting point, let us assume a generic $(d+1)$-dimensional supergravity low-energy effective action of the form
\begin{align}\label{eq:sugra_action}
	\S_{\text{eff}} = \eta \int_{\text{AdS}_{d+1}} d^{d+1}x\sqrt{g}\,\big(-R+\L_{\text{eff}}(\Phi_i)\big),
\end{align}
for some effective Lagrangian $\L_{\text{eff}}$ depending on a collection of scalar fields $\Phi_i$. The overall factor $\eta$ determines the normalization of the action and in top-down AdS/CFT constructions is determined by the parent higher-dimensional string or supergravity theory. In the top-down model of interest in this work $\eta$ is given in~\eqref{eq:S_eff}. As discussed in the previous section, one can perform field redefinitions of the bulk fields such that they have canonically normalized and diagonal kinetic and mass terms. The bulk Lagrangian, expanded in small values of the fields, then takes the schematic form\footnote{From here on we sometimes set the AdS$_5$ length scale $L=1$ to simplify some of the expressions below and conform to the existing literature.} 
\begin{align}\label{eq:Leff}
\begin{split}
	\mathcal{L}_{\text{eff}} &= \frac{1}{2}(\nabla_{\mu}\Phi_i)^2 + d_{ijk}\Phi_i(\nabla_{\mu}\Phi_j)(\nabla^{\mu}\Phi_k)+\Lkin^{(4)}+\ldots\\
	&\quad+\frac{1}{L^2}\Big[-d(d-1)+m_i^2\Phi_i^2+c_{ijk}\Phi_i\Phi_j\Phi_k+\P^{(4)}+\ldots\Big].
\end{split}
\end{align}

First, we discuss the 2pt-functions. Having diagonal kinetic terms in \eqref{eq:Leff} leads to the 2pt-functions being diagonal as well. To fix the normalization one needs to carefully solve the wave-equation for a massive scalar $\Phi$ in AdS (with dual operator $\widehat{\o}_\Phi$) to find
\begin{align}\label{eq:normalised_2pt}
	\langle\widehat{\o}_{\Phi}(x_1)\widehat{\o}_{\Phi}(x_2)\rangle = \frac{n_{\Delta}}{\abs{x_1-x_2}^{2\Delta}}\,,\quad\text{with}\quad n_\Delta=\eta\,\frac{(2\Delta-d)\Gamma(\Delta)}{\pi^{d/2}\Gamma(\Delta-\frac{d}{2})}\,,
\end{align}
where $\Delta$ is the conformal dimension of $\widehat{\o}_{\Phi}$ determined by the mass $m$ of the scalar field through the standard AdS/CFT relation
\begin{equation}
m^2L^2 = \Delta(\Delta-d)\,.
\end{equation}
We are ultimately interested in correlation functions of \textit{normalised} operators, henceforth denoted by $\o_{\Phi}$, which is achieved by the rescaling\footnote{Note that the 2pt-function normalisation $n_\Delta$ has a double zero for $\Delta=\frac{d}{2}$. In that case, the formula \eqref{eq:normalised_2pt} for $n_\Delta$ should be modified, see the discussion around eq. (2.21) of \cite{Klebanov:1999tb}. For the 3pt-function however, there is no such subtlety: normalising the 3pt-function according to \eqref{eq:normalised_operator} the zero of $n_\Delta$ at $\Delta=\frac{d}{2}$ cancels.}
\begin{align}\label{eq:normalised_operator}
	\o_{\Phi}:=\frac{\widehat{\o}_{\Phi}}{\sqrt{n_{\Delta}}}\,.
\end{align}
From now on, we will take all operators $\o_{\Phi_i}$ to be unit normalised by means of \eqref{eq:normalised_2pt}-\eqref{eq:normalised_operator}.

\begin{figure}[t]
\begin{center}
\begin{tikzpicture}[scale=0.5]
\begin{scope}
\draw (0, 0) circle (3);
\node[circle, fill=black, inner sep=1pt] (V) at (0,0) {};
\node[coordinate, label=above:$c_{ijk}$] at (0.5,-0.1) {};
\node[coordinate, label=right:$\Phi_j$] (x1) at (3,0) {};
\node[coordinate, label=below:$\Phi_k$] (x2) at (-1.5,-2.6) {};
\node[coordinate, label=above:$\Phi_i$] (x3) at (-1.5,2.6) {};
\draw (V) -- (x1);
\draw (V) -- (x2);
\draw (V) -- (x3);
\end{scope}
\begin{scope}[xshift=11cm]
\draw (0, 0) circle (3);
\node[circle, fill=black, inner sep=1pt] (V) at (0,0) {};
\node[coordinate, label=above:$d_{ijk}$] at (0.5,-0.1) {};
\node[coordinate, label=right:$\Phi_j$] (x1) at (3,0) {};
\node[coordinate, label=below:$\Phi_k$] (x2) at (-1.5,-2.6) {};
\node[coordinate, label=above:$\Phi_i$] (x3) at (-1.5,2.6) {};
\draw (V) -- (x1);
\draw (V) -- (x2);
\draw (V) -- (x3);
\node at (1.5, -0.5) {$\partial$};
\node at (-0.3,-1.3) {$\partial$};
\end{scope}
\end{tikzpicture}
\caption{3pt-Witten diagrams. On the left: non-derivative cubic vertex $c_{ijk}$. On the right: cubic vertex $d_{ijk}$ with derivatives on $\Phi_j$ and $\Phi_k$.}
\label{fig:3pt}
\end{center}
\end{figure}
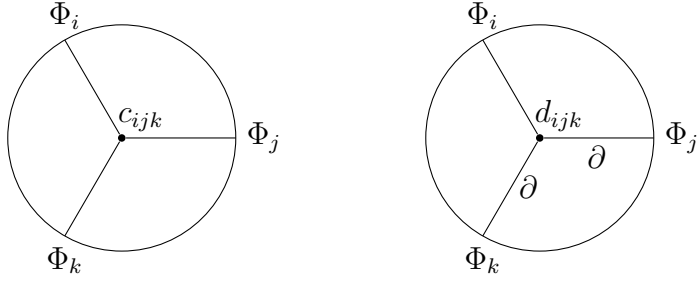

The 3pt-functions are computed by evaluating the Witten diagrams depicted in Figure~\ref{fig:3pt}. These diagrams can be evaluated using the method of \cite{Freedman:1998tz} and the resulting expression takes the expected form of a CFT 3pt-function given by
\begin{align}\label{eq:3pt_func}
	\langle\o_{\Phi_i}(x_1)\o_{\Phi_j}(x_2)\o_{\Phi_k}(x_3)\rangle = \frac{C_{\Phi_i\Phi_j\Phi_k}}{\abs{x_1-x_2}^{\Delta_1+\Delta_2-\Delta_3}\abs{x_1-x_3}^{\Delta_1+\Delta_3-\Delta_2}\abs{x_2-x_3}^{\Delta_2+\Delta_3-\Delta_1}}\,.
\end{align}
The 3pt-function coefficient $C_{\Phi_i\Phi_j\Phi_k}$ is given in terms of the supergravity cubic couplings $c_{ijk}$ and $d_{ijk}$, which are read off from \eqref{eq:Leff}:
\begin{empheq}[box=\fbox]{equation}
\begin{aligned}\label{eq:C_3pt}
	~C_{\Phi_i\Phi_j\Phi_k} = c_{ijk}\,A_1 + d_{ijk}\,A_2\,,~
\end{aligned}
\end{empheq}
where the $A_i$ are functions of the dimensions $\Delta_i$ of the external operators $\o_{\Phi_i}$. For unit normalised operators, one has
\begin{align}\label{eq:norm_3pt}
\begin{split}
	A_1 &= \frac{1}{\sqrt{\eta}}\,\frac{\Gamma\big(\frac{\Delta_i+\Delta_j-\Delta_k}{2}\big)\Gamma\big(\frac{\Delta_i+\Delta_k-\Delta_j}{2}\big)\Gamma\big(\frac{\Delta_j+\Delta_k-\Delta_i}{2}\big)}{4\pi^{d/4}\sqrt{2\prod_{n\in\{i,j,k\}}^3\Gamma\big(\Delta_n\big)\Gamma\big(\Delta_n+1-\frac{d}{2}\big)}}\,\Gamma\Big(\frac{\Delta_i+\Delta_j+\Delta_k-d}{2}\Big)\,,\\[5pt]
	A_2 &=\big[\Delta_j\Delta_k+\tfrac{1}{2}(d-\Delta_i-\Delta_j-\Delta_k)(\Delta_j+\Delta_k-\Delta_i)\big]\,A_1\,.
\end{split}
\end{align}
Note that the overall normalization $\eta$ of the gravitational effective action \eqref{eq:sugra_action} enters as a factor of $1/\sqrt{\eta}$ in the 3pt-functions.

Note that when reading off the cubic couplings $c_{ijk}$ and $d_{ijk}$ from the bulk Lagrangian one needs to be careful with \textit{symmetry factors}. Indeed, in \eqref{eq:Leff} we assumed that the three bulk fields which enter the cubic terms are all distinct from each other. If two of them or all three fields are equal, one must include an appropriate symmetry factor in front of the cubic coupling to account for the different Wick contractions. This results in an extra factor of $2!$ for 3pt-function coefficients involving two identical operators, $C_{\Phi_i\Phi_i\Phi_j}$, and a factor of $3!$ for three identical operators, $C_{\Phi_i\Phi_i\Phi_i}$.

\subsection{Removal of all derivative-couplings}
\label{subsec:removal}
While the previously described procedure is fully general, there is a simplification of the cubic couplings which is worth discussing. It turns out that one can trade all derivative-couplings $d_{ijk}$ for polynomial non-derivative couplings $c_{ijk}$, and thus for the purposes of calculating holographic 3pt-functions effectively remove all $d_{ijk}$-terms from the effective Lagrangian. The mechanism that realizes this is very simple, as we now describe. Consider the following total derivative: 
\begin{align}\label{eq:total_derivative}
	\nabla_{\mu}\big(\Phi_i\Phi_j\nabla^{\mu}\Phi_k+\Phi_i\Phi_k\nabla^{\mu}\Phi_j-\Phi_j\Phi_k\nabla^{\mu}\Phi_i\big) = 2\Phi_i\nabla_{\mu}\Phi_j\nabla^{\mu}\Phi_k - (m_i^2-m_j^2-m_k^2)\Phi_i\Phi_j\Phi_k\,.
\end{align}
To obtain the terms involving the masses $m_i^2$ on the right hand side we have used the quadratic bulk equation of motion for the scalars, $(-\nabla^2+m_i^2)\Phi_i=0$. This operation is legitimate if one is interested in the calculation of holographic 3pt-functions since the bulk scalar fields need to be on-shell. Additionally, it is enough to restrict to the linearized equations of motion since the fields $\Phi_i$ are small excitations near the AdS vacuum of interest. Since the term on the left hand side of \eqref{eq:total_derivative} is a total derivative, we can add it with an arbitrary constant coefficient to the bulk effective Lagrangian.\footnote{In this discussion we ignore the possibility that there are finite boundary counterterms that should be used in the holographic renormalization procedure when computing holographic correlators. In some concrete examples of AdS$_4$/CFT$_3$ such terms are known to arise and play a crucial role in the proper calculation of holographic correlators, see \cite{Freedman:2013oja,Freedman:2016yue,Bobev:2018uxk,Bobev:2018wbt} as well as the discuss in Section~\ref{subsec:bdry} below.}
We therefore arrive at the following new effective Lagrangian which is equivalent for the purpose of calculating 3pt holographic correlators:
\begin{align}\label{eq:Leff_shift}
	\L_{\text{eff}}'~\widehat{=}~\L_{\text{eff}}+\gamma_{ijk}\,\Big(-\Phi_i\nabla_{\mu}\Phi_j\nabla^{\mu}\Phi_k+\frac{m_i^2-m_j^2-m_k^2}{2}\,\Phi_i\Phi_j\Phi_k\Big)\,.
\end{align}
Here $\gamma_{ijk}$ are free parameters which we can tune to any desired value. In particular, by choosing $\gamma_{ijk}=d_{ijk}$ we can replace any cubic derivative-term $\Phi_i\nabla_{\mu}\Phi_j\nabla^{\mu}\Phi_k$ by a corresponding term \textit{without} derivatives. In other words, at the level of cubic terms in the Lagrangian we effectively have the map
\begin{empheq}[box=\fbox]{equation}
\begin{aligned}\label{eq:d_to_c}
	~\Phi_i\nabla_{\mu}\Phi_j\nabla^{\mu}\Phi_k \mapsto \frac{m_i^2-m_j^2-m_k^2}{2}\,\Phi_i\Phi_j\Phi_k\,.~
\end{aligned}
\end{empheq}
It is important to note that this relation is consistent with the previously described holographic computation of 3pt-functions, in particular with equations \eqref{eq:C_3pt} and \eqref{eq:norm_3pt}. Indeed, upon using the relation $m_i^2=\Delta_i(\Delta_i-d)$ for each scalar $\Phi_i$, one can check that the conversion factor $\frac{1}{2}(m_i^2-m_j^2-m_k^2)$ precisely equals the factor relating $A_1$ to $A_2$ in the second line of \eqref{eq:norm_3pt}.

Lastly, let us emphasize that the relation \eqref{eq:d_to_c} is \textit{not} a field redefinition. All we have done is a rewriting of the cubic terms alone, and as such all higher-order terms in the effective Lagrangian remain unaffected. In the process of deriving \eqref{eq:d_to_c} we have used the linearized equations of motion which is allowed since we are interested in the calculation of holographic 3pt-functions. As explained in previous literature, see e.g. \cite{Liu:1998th}, the same effect of removing the cubic derivative-couplings can be achieved by performing a quadratic field redefinition. The two procedures are related but not equivalent. The field redefinition discussed in \cite{Liu:1998th} is ``off-shell'' and may be permissible when $\L_{\text{eff}}$ has only polynomial terms in the scalar field potential, but in top-down holographic models this is typically not the case. As discussed in Section~\ref{sec:10-scalar}, type IIB supergravity and its 5d gauged supergravity truncation lead to a fully non-linear action in which the potential is not a polynomial in the scalar fields. Performing a quadratic field redefinition to such an action introduces new higher-order couplings which may have unintended consequences for the calculation of higher-point functions. This potential issue can be avoided by working at a fixed order in the expansion around the vacuum solution, and the rewriting \eqref{eq:d_to_c} is an explicit example of this at cubic order relevant for the calculation of 3pt-functions. Similar methods may be adapted to the holographic calculation of higher-point functions.

\subsection{Extremal 3pt-functions and single-particle operators}
\label{subsec:extremal_couplings}

Studying the relation between bulk cubic couplings and the dual 3pt-function coefficient, an alert reader may be worried about possible divergences in the gamma functions in~\eqref{eq:norm_3pt}. Indeed, the 3pt-Witten diagram is divergent in the case of a so-called extremal configuration of operator dimensions: the coefficients $A_1$ and $A_2$ develop a simple pole when $\Delta_i=\Delta_j+\Delta_k$ or permutations thereof.\footnote{Another potential source for a divergence is the fourth gamma function in $A_1$, c.f. \eqref{eq:norm_3pt}, which diverges when $\Delta_i+\Delta_j+\Delta_k=d$. This configuration is only possible for operator dimensions very close to the unitarity bound. We will not consider this somewhat exotic case here as it is not realised in the holographic setup we study. However, we note in passing that in the ABJM 3d SCFT there are protected operators with $\Delta=1$ which realise this scenario and need to be treated carefully in the bulk. See \cite{Freedman:2016yue} for a detailed discussion of the holographic 3pt-functions for these operators.}
Note that this situation is ubiquitous in (half-)maximally supersymmetric examples of holography due to the presence of infinite KK-towers of chiral operators with integer dimensions, resulting in an \`a priori infinite set of extremal couplings. In the 4d $\N=1$ LS SCFT of interest here extremal arrangements of operator dimensions occur, even in the 10-scalar truncation introduced in Section \ref{sec:10-scalar}, c.f. Table \ref{tab:PW_spectrum}.\footnote{This appears to also be the case in other 4d $\N=1$ holographic SCFTs arising from D3-branes. For instance, for the Klebanov-Witten SCFT one can deduce that extremal arrangements of operator dimensions exist by inspecting the KK tower of supergravity modes in the dual  AdS$_5\times T^{1,1}$ background, see \cite{Ceresole:1999zs}.} 

Since the correlation functions in the boundary CFT are expected to be finite, the poles in $A_1$ and $A_2$ arising from the extremal arrangements of operator dimensions must somehow get cancelled.\footnote{See \cite{Korchemsky:2015cyx} for an interesting example where a seemingly diverging extremal 3pt-function in the planar limit of $\mathcal{N}=4$ SYM becomes finite when $1/N$ effects are taken into account.} Generically, there are three possibilities for how such a cancellation for extremal cubic couplings $c_{ijk}$ and $d_{ijk}$ might occur:
\begin{enumerate}[label=(\arabic{enumi})]
\item $c_{ijk}=d_{ijk}=0$. The simplest resolution is when both cubic couplings vanish separately, providing the necessary zero to cancel the divergence. This is in fact the default situation realised in all known maximally supersymmetric setups.\footnote{For supergravity on AdS$_5\times$S$^5$, the vanishing of extremal cubic couplings was noted in \cite{DHoker:1998ecp,Arutyunov:1999en}, the relation of this fact to the existence of 5d supergravity consistent truncations was emphasized in \cite{DHoker:2000pvz} and more recently derived using ExFT methods in~\cite{Duboeuf:2023cth}.} 
\item $c_{ijk}=0$, $d_{ijk}\neq0$. This case can occur if the prefactor in $A_2$ provides the necessary zero, i.e. for an extremal configuration of dimensions which furthermore obey
\begin{align} \Delta_j\Delta_k+\tfrac{1}{2}(d-\Delta_i-\Delta_j-\Delta_k)(\Delta_j+\Delta_k-\Delta_i)=0\,.
\end{align}
\item $c_{ijk}\neq0$, $d_{ijk}\neq0$. In this situation, the divergences in $A_1$ and $A_2$ must non-trivially cancel in the combination $c_{ijk}A_1+d_{ijk}A_2$ so as to yield a finite 3pt-function coefficient in the dual CFT. For this to happen, the bulk cubic couplings must be fine-tuned to satisfy one of the following conditions:
\begin{equation}
  \begin{alignedat}{2}
    c_{ijk}&=-\Delta_j\Delta_k\,d_{ijk}\,, &\qquad\text{for }\Delta_i=\Delta_j+\Delta_k\,,\\[2pt]
    c_{ijk}&=-(d-\Delta_j)\Delta_k\,d_{ijk}\,, &\text{for }\Delta_j=\Delta_i+\Delta_k\,,\\[2pt]
    c_{ijk}&=-(d-\Delta_k)\Delta_j\,d_{ijk}\,, &\text{for }\Delta_k=\Delta_i+\Delta_j\,.
  \end{alignedat}
\end{equation}
\end{enumerate}
The situations in (2) and (3) above disagree with the common lore that extremal cubic couplings need to vanish in AdS/CFT. Perhaps the reason why these possibilities have been overlooked in the past is due to the focus on maximally supersymmetric setups, where case (1) indeed applies. In more general top-down holographic setups, like the one we consider in this work, it is not clear why situations (2) and (3) cannot arise. Indeed, as we show explicitly in Section~\ref{sec:3pt_IR}, all three options discussed above are realized in the 10-scalar model cubic couplings around the IR AdS$_5$ vacuum. Interestingly, the fine-tuning required for cases (2) and (3) to work can be combined with the mechanism discussed in Section~\ref{subsec:removal} to convert (2) or (3) to (1) by an addition of an appropriate total derivative term. It is therefore always possible to rewrite the supergravity cubic terms in a convenient `frame' where all extremal cubic couplings vanish. Whether this is the natural formulation arising from a particular supergravity effective action or consistent truncation depends on the given top-down holographic construction.  

The discussion above was focussed on the conditions under which extremal couplings in the bulk supergravity action result in finite 3pt-functions. We now turn to the dual CFT perspective, and in particular we address the question of which finite value one should assign to a given extremal holographic 3pt-function. This is in fact a rather subtle question, which has led to some confusion in the past.\footnote{See \cite{Castro:2024cmf} for a recent discussion on an alternative bottom-up approach to this subtlety.} After all, in cases (1) and (2) above the naive holographic prescription produces a result of the form $0\cdot\infty$, to which one may in principle assign any desired value for the dual CFT 3pt-function coefficient.\footnote{In the above case (3), one may argue that the finite piece of the sum $c_{ijk}A_1+d_{ijk}A_2$ gives the ``correct'' CFT 3pt-function coefficient. However, this is an artefact of the specific parametrisation of the supergravity cubic couplings. By using an arbitrary fraction of the total derivative shift in \eqref{eq:Leff_shift}, one can obtain new cubic couplings $c'_{ijk}$ and $d'_{ijk}$ from which any desired non-zero value for the CFT extremal 3pt-function may be achieved.}

It turns out that this question can not be answered directly from the two-derivative bulk supergravity point of view. Instead, the apparent arbitrariness of extremal 3pt-functions is related to a subtlety in the correspondence between bulk fields $\Phi$ and their dual CFT operators $\o_{\Phi}$. More precisely, for every extremal arrangement of operator dimensions there is an ambiguity in the field-operator holographic dictionary due to mixing between single- and multi-trace operators. This ambiguity was pointed out in early work on AdS/CFT, see \cite{Liu:1999kg,DHoker:1999jke,Arutyunov:2000ima}, and we discuss it in some detail below, together with a proposal of how to remove it.

To keep the discussion general we will not assume any microscopic top-down realization of AdS/CFT. To distinguish between single and double-trace operators in the large-$N$ limit of the correspondence we can consider the scaling of correlation functions in the CFT with the gravitational coupling $\eta$ in \eqref{eq:sugra_action} that controls the large-$N$ expansion. In general $\eta$ scales as a positive power of $N$ which is determined by the concrete holographic model. In our setup arising from $N$ D3-branes in type IIB supergravity we have $\eta \sim N^2$, see~\eqref{eq:S_eff}. We now consider 2pt- and 3pt-functions of single trace, $\mathcal{O}_i$, and double-trace, $\mathcal{O}_i^2$, operators in this holographic model. These correlators have the following large $\eta$ scaling 
\begin{equation}\label{eq:mtracelargeN}
\langle \mathcal{O}_i\mathcal{O}_i\rangle \sim \eta^0\,, \qquad \langle \mathcal{O}_i^2\mathcal{O}_i^2\rangle \sim \eta^{0}\,, \qquad \langle \mathcal{O}_i\mathcal{O}_j\mathcal{O}_k\rangle \sim \eta^{-1/2}\,, \qquad \langle \mathcal{O}_i^2\mathcal{O}_j\mathcal{O}_k\rangle \sim \eta^{0}\,.
\end{equation}
Importantly, to derive the last scaling we have assumed that the double-trace operator $\mathcal{O}_i^2$ is obtained by taking the coincident point limit of the single trace operators $\mathcal{O}_j$ and $\mathcal{O}_k$, i.e. we have $\Delta[\mathcal{O}_i^2] = \Delta[\mathcal{O}_j]+\Delta[\mathcal{O}_k]$. Equipped with this large $\eta$ scaling we then can unambiguously distinguish single-trace and double-trace operators in this somewhat abstractly defined large $\eta$ CFT.

We now consider a scalar operator $\o_{\Phi}$ of dimension $\Delta$ in the CFT which allows for an extremal configuration with two other operators of dimension $\Delta_1$ and $\Delta_2$, i.e. we have that $\Delta=\Delta_1+\Delta_2$. The composite `double-trace' operator $\o_{\Phi_1,\Phi_2}\equiv[\o_{\Phi_1}\o_{\Phi_2}]$, defined by the coincidence limit of $\o_{\Phi_1}$ and $\o_{\Phi_2}$, then also has dimension $\Delta$ by the assumption of extremality and the properties of the OPE in CFTs. We therefore have a degeneracy in the CFT spectrum and the operators $\o_{\Phi}$ and $\o_{\Phi_1,\Phi_2}$ will generically mix. In the context of holography this leads to an ambiguity in the holographic dictionary stemming from the fact that we cannot immediately identify the linear combination of operators dual to the gravitational scalar field $\Phi$ with mass $m^2L^2 = \Delta(\Delta-d)$. We then conclude that the operator dual to the bulk field $\Phi$ is not necessarily the `single-trace' operator $\o_{\Phi}$, but a linear combination of the form 
\begin{align}\label{eq:admixture}
	\Phi\quad\leftrightarrow\quad\o'_{\Phi}=\o_{\Phi}+\mathfrak{c}\,\o_{\Phi_1,\Phi_2}+\ldots\,.
\end{align}
The coefficient $\mathfrak{c}$ is a real number that parametrizes the ambiguity due to the operator mixing. The ellipsis represent possible $k$-trace admixtures, with $k>2$, arising from additional operator degeneracies from CFT operators in the spectrum that obey $\Delta=\Delta_{1}+\ldots+\Delta_{k}$. From this point of view, the ambiguity of calculating holographic extremal 3pt-functions manifested by the freedom to change the parametrisation of supergravity couplings by means of \eqref{eq:Leff_shift}, amounts to changing the value of the coefficient $\mathfrak{c}$ in \eqref{eq:admixture}. The arbitrariness of extremal 3pt-functions is thus related to the precise choice of operator dual to the bulk field $\Phi$. Note that this ambiguity only affects bulk fields whose dimension allows for an extremal configuration. For all other fields in the bulk theory there is no ambiguity in the identification between bulk field and CFT operators, unless the spectrum is degenerate for other reasons. Notice that due to the large $\eta$ scaling in \eqref{eq:mtracelargeN} the coefficient $\mathfrak{c}$ must scale as $\mathfrak{c} \sim \eta^{-1/2}$ and therefore the operator mixing is suppressed in the large $\eta$ limit. The coefficients of any possible multi-trace admixtures in the ellipsis in \eqref{eq:admixture} are even further suppressed by higher powers of $1/\eta$.

So, how do we fix the correct linear combination in \eqref{eq:admixture} in holographic CFTs? A compellingly simple answer to this question has been formulated in the context of the $\N=4$ SYM theory and its holographic dual in \cite{Aprile:2018efk,Aprile:2019rep},\footnote{The possibility of having admixtures of single- and multi-trace operators was discussed in the early days of AdS/CFT \cite{Arutyunov:1999en,DHoker:1999jke}, precisely in the context of extremal 3pt-functions. However, only later on in the study of 4pt-functions of higher-KK modes at tree-level the necessity of double-trace admixtures became apparent \cite{Arutyunov:2018neq}. The correct identification of the dual operators has then been of crucial importance for computing loop corrections to holographic 4pt-functions, see \cite{Aprile:2019rep} and \cite{Alday:2019nin}. Moreover, similar operator mixing has been discussed for the supergravity spectrum on AdS$_3\times$S$^3$, see \cite{Taylor:2007hs} and \cite{Rawash:2021pik}, and in one-loop corrections to supergluon scattering on AdS$_5\times$S$^3$ \cite{Huang:2023ppy}.}
which is readily applicable to any holographic setup: 
\vspace{-0.2cm}
\setlength{\fboxsep}{7pt}
\begin{center}
\noindent\fbox{\begin{minipage}{0.7\textwidth}
	The operator dual to a single-particle state in the bulk is the unique operator which is \textit{orthogonal to all multi-trace operators}.
\end{minipage}}
\end{center}\vspace{-0.2cm}
This prescription singles out a precise linear combination of multi-trace admixtures in \eqref{eq:admixture}, defining the so-called single-particle operator (SPO) \cite{Aprile:2019rep} which we will denote by $\o_{\Phi}^{(s)}$. The statement above can be translated into a vanishing condition on the two point function of the SPO $\o_{\Phi}^{(s)}$ with all multi-trace operators in the CFT, i.e. 
\begin{align}\label{eq:orthogonality}
	\langle\,\o_{\Phi}^{(s)}(x)\,[\o_{\Phi_{1}}\cdots\o_{\Phi_{k}}](y)\,\rangle=0\,, \qquad k\geq 2\,.
\end{align}
From the perspective of the bulk gravitational action the prescription above ensures that the bulk field sourced by $\o_{\Phi}^{(s)}$ is orthogonal to all multi-particle states. Generically, requiring \eqref{eq:orthogonality} will fix the coefficient $\mathfrak{c}$, as well as the coefficients of the omitted higher-trace terms, in \eqref{eq:admixture} to some non-zero value. We thus conclude that a basic tenet of the AdS/CFT correspondence, namely the statement that bulk fields correspond to single-trace operators in the boundary CFT, needs to be modified when extremal arrangements of operator dimensions are present in the spectrum. Indeed, bulk fields with dimensions allowing extremal configurations are generically dual to a linear combination of single- and multi-trace operators.  Of course, from a practical point of view it may be very hard to compute all necessary 2pt-functions in a given holographic CFT to solve the operator mixing problem outlined above and find the explicit value of $\mathfrak{c}$. However, in the prime example of $\N=4$ SYM we are in a situation where the operator dimensions corresponding to bulk KK supergravity modes are protected and the theory admits a weakly coupled point on the conformal manifold. This, together with the definition \eqref{eq:orthogonality}, was used in \cite{Aprile:2020uxk} to explicitly construct the SPO of any dimension and at finite $N$.

In summary, the apparent arbitrariness of extremal 3pt-functions is related to operator mixing in the CFT. However, a natural choice of the operator dual to a single-particle state in the bulk is the SPO, defined as in \eqref{eq:orthogonality} to be orthogonal to all multi-particle states. In turn, the orthogonality relation \eqref{eq:orthogonality}, which is valid for arbitrary values of $\eta$ (or $N$), is equivalent to fixing the extremal 3pt-function to vanish. We therefore arrive at the following proposal for fixing the ambiguity in the calculation of extremal holographic 3pt-functions:
\vspace{-0.2cm}
\setlength{\fboxsep}{7pt}
\begin{center}
\noindent\fbox{\begin{minipage}{0.82\textwidth}
	The extremal holographic 3pt-functions of single-particle operators vanish.
\end{minipage}}
\end{center}\vspace{-0.2cm}

\section{Holographic 3pt-functions for the LS SCFT}
\label{sec:3pt_IR}
We now return to our concrete supergravity effective action, namely the 10-scalar model introduced in Section~\ref{sec:10-scalar}, and apply the general procedure of computing holographic 3pt-functions from Section~\ref{sec:3pt_general}. Here we will focus on the holographic calculation of 3pt-functions of the theory in the IR, i.e. the LS SCFT. The corresponding computation for the maximally supersymmetric solution in the UV dual to the $\mathcal{N}=4$ SYM theory is performed in Appendix~\ref{app:3pt_UV}. There we also show agreement with 3pt-functions computed in $\N=4$ SYM theory, thus providing a consistency check of the method.

\subsection{Bulk cubic couplings in the IR}
\label{subsec:cubic_IR}
The starting point for the computation of 3pt-functions are the cubic terms in the effective Lagrangian. We will first present the cubic terms denoted by $\Lkin^{(3)}$ and $\P^{(3)}$ in \eqref{eq:L_IR_exp} which originate from expanding the fully non-linear action \eqref{eq:L_full} of the 10-scalar model. In the notation introduced in the previous section, this yields a set of cubic couplings $d_{ijk}$ and $c_{ijk}$. These couplings are directly obtained from supergravity and hence define what we will call the `original frame'  or frame 1. However, as discussed in Section~\ref{subsec:removal}, it is possible to remove all derivative-couplings $d_{ijk}$ and turn them into non-derivative ones. We will refer to this simpler set of cubic couplings, now consisting only of $c_{ijk}$ terms, as frame 2.

\subsubsection{Cubic couplings from supergravity: frame 1}
\label{subsec:frame1}
Performing the expansion around the IR AdS$_5$ solution to cubic order and implementing the three-step procedure \eqref{eq:step1}-\eqref{eq:step3}, we find the following cubic couplings. From the potential, we obtain
\begin{align}\label{eq:P3_IR}
\begin{split}
	\mathcal{P}^{(3)} &= \frac{9}{8}\big(2\alpha_1\alpha_2\phi_2+2\alpha_1\alpha_3\phi_3-\alpha_2\alpha_3\phi_4\big)-\frac{21}{8}\phi_2\phi_3\phi_4 \\
	&\quad+\frac{3}{4\sqrt{2}}\,\bh\,\big(\alpha_2^2-\alpha_3^2-\phi_2^2+\phi_3^2\big)\\
	&\quad+\frac{\sqrt{3}}{16\sqrt{14}}\big(\alpha_2^2+\alpha_3^2\big)\Big(\sqrt{917+29 \sqrt{7}}\,\rho_1-\sqrt{917-29 \sqrt{7}}\,\rho_2\Big)\\
	&\quad-\frac{\sqrt{3}}{16\sqrt{2}}\big(\phi_2^2+\phi_3^2\big)\Big(\sqrt{371+107 \sqrt{7}}\,\rho_1+\sqrt{371-107 \sqrt{7}}\,\rho_2\Big)\\	&\quad+\frac{1}{2\sqrt{21}}\,\bh^2\Big(\sqrt{217+79\sqrt{7}}\,\rho_1-\sqrt{217-79 \sqrt{7}}\,\rho_2\Big)\\
	&\quad-\frac{3}{56}\,\phi_4^2 \Big(\sqrt{686+238\sqrt{7}}\,\rho_1-\sqrt{686-238 \sqrt{7}}\,\rho_2\Big)\\
	&\quad+\frac{\sqrt{3}}{2\sqrt{14}}\,\rho_1\rho_2\Big(\sqrt{35+13\sqrt{7}}\,\rho_1-\sqrt{35-13\sqrt{7}}\,\rho_2\Big)\\
	&\quad-\frac{\sqrt{3}}{4\sqrt{14}}\,\alpha_1^2\Big(\sqrt{2891-517\sqrt{7}}\,\rho_1-\sqrt{2891+517\sqrt{7}}\,\rho_2\Big)\\
	&\quad+\frac{1}{6\sqrt{42}}\Big(\sqrt{36155-13261\sqrt{7}}\,\rho_1^3+\sqrt{36155+13261\sqrt{7}}\,\rho_2^3\Big)\,. 
\end{split}
\end{align}
On the other hand, the cubic terms coming from expanding the kinetic term are given by
\begin{align}\label{eq:L3_IR}
\begin{split}
	\mathcal{L}_{\text{kin}}^{(3)} &=-\frac{1}{2}\big(\phi_2\parm\alpha_1\parmm\alpha_2+\phi_3\parm\alpha_1\parmm\alpha_3-\phi_4\parm\alpha_2\parmm\alpha_3\big)\\
	&\quad-\frac{1}{2}\big(\phi_2\parm\alpha_3+\phi_3\parm\alpha_2-\phi_4\parm\alpha_1\big)\parmm\varphi\\
	&\quad+\frac{1}{2}\big(\phi_2\parm\phi_3\parmm\phi_4+\phi_3\parm\phi_2\parmm\phi_4+\phi_4\parm\phi_2\parmm\phi_3\big)\\
	&\quad+\frac{1}{4\sqrt{14}}\Big(\sqrt{7-\sqrt{7}}\,\rho_1+\sqrt{7+\sqrt{7}}\,\rho_2\Big)\\
	&\quad\quad\times\Big[(\parm\alpha_1)^2+(\parm\alpha_2)^2+(\parm\alpha_3)^2+(\parm\phi_2)^2+(\parm\phi_3)^2+(\parm\varphi)^2+(\parm\phi_4)^2\Big]\\
	&\quad+\frac{1}{2\sqrt{14}} \big(\phi_2\parm\phi_2+\phi_3\parm\phi_3+\phi_4\parm\phi_4\big)\Big(\sqrt{7-\sqrt{7}}\,\parmm\rho_1+\sqrt{7+\sqrt{7}}\,\parmm\rho_2\Big)\\
	&\quad+\frac{\sqrt{3}}{28}\Big(\sqrt{7-\sqrt{7}}\,\rho_1+\sqrt{7+\sqrt{7}}\,\rho_2\Big)\parm\rho_1\parmm\rho_2\\
	&\quad+\frac{1}{56}\rho_1\Big(\sqrt{35-11\sqrt{7}}\,(\parm\rho_1)^2 +\sqrt{21+3\sqrt{7}}\,(\parm\rho_2)^2\Big)\\
	&\quad+\frac{1}{56}\rho_2\Big(\sqrt{21-3 \sqrt{7}}\,(\parm\rho_1)^2+\sqrt{35+11\sqrt{7}}\,(\parm\rho_2)^2\Big)\,.
\end{split}
\end{align}
This summarises the explicit cubic couplings in the original frame, as derived from the non-linear 5d $\mathcal{N}=8$ gauged supergravity Lagrangian. As we now discuss, this may not be the optimal parametrisation in the sense that the above cubic couplings can be greatly simplified.

\subsubsection{Removal of all derivative-couplings: frame 2}
\label{subsec:frame2}
We now implement the mechanism described in Section~\ref{subsec:removal}, allowing us to move to a new frame of supergravity couplings where all $d_{ijk}$-type terms vanish. The key relation is \eqref{eq:d_to_c}, which converts derivative-couplings into non-derivative ones. It turns out that the new terms combine nicely with the $c_{ijk}$ couplings in \eqref{eq:P3_IR}, defining what we call frame 2. In this alternative frame we thus have $\Lkin^{(3)}\vert_{\text{frame 2}}=0$, while the non-vanishing cubic couplings take the form
\begin{align}\label{eq:P3_IR_new}
\begin{split}
	\mathcal{P}^{(3)}\vert_{\text{frame 2}} &= \frac{3}{2}\alpha_1(2\alpha_2\phi_2+2\alpha_3\phi_3-\varphi\phi_4)+\frac{3}{4\sqrt{2}}\,\bh\,\big(\alpha_2^2-\alpha_3^2-\phi_2^2+\phi_3^2\big)\\
	&\quad+\frac{\sqrt{3}}{4}\big(\alpha_2^2+\alpha_3^2-\phi_2^2-\phi_3^2\big)\Big(\sqrt{7+\sqrt{7}}\,\rho_1+\sqrt{7-\sqrt{7}}\,\rho_2\Big)\\
	&\quad+\frac{1}{2\sqrt{21}}\,\bh^2\Big(\sqrt{217+79\sqrt{7}}\,\rho_1-\sqrt{217-79\sqrt{7}}\,\rho_2\Big)\\
	&\quad-\frac{\sqrt{3}}{4\sqrt{14}}\,\phi_4^2\Big(\sqrt{77+29\sqrt{7}}\,\rho_1-\sqrt{77-29\sqrt{7}}\,\rho_2\Big)\\
	&\quad-\frac{\sqrt{3}}{4\sqrt{14}}\,\varphi^2\Big(\sqrt{35-13\sqrt{7}}\,\rho_1-\sqrt{35+13\sqrt{7}}\,\rho_2\Big)\\
	&\quad+\frac{\sqrt{3}}{56}\,\rho_1\rho_2\Big(\sqrt{2065+377\sqrt{7}}\,\rho_1-\sqrt{2065-377\sqrt{7}}\,\rho_2\Big)\\
	&\quad-\frac{\sqrt{3}}{4\sqrt{14}}\,\alpha_1^2\Big(\sqrt{3296-523\sqrt{7}}\,\rho_1-\sqrt{3296+523\sqrt{7}}\,\rho_2\Big)\\
	&\quad+\frac{1}{168\sqrt{3}}\Big(\sqrt{1659343-602999\sqrt{7}}\,\rho_1^3+\sqrt{1659343+602999\sqrt{7}}\,\rho_2^3\Big)\,.
\end{split}
\end{align}
Note that this rewriting is more compact than the couplings in the original frame 1, c.f. \eqref{eq:P3_IR}-\eqref{eq:L3_IR}. Despite having fewer terms, when it comes to the calculation of holographic 3pt-functions, the same information is contained in frame 2.

Upon a closer look at the cubic couplings one notices a few subtle differences. Certain cubic couplings which were present in frame 1 vanish in frame 2. This is the case for couplings involving the fields $\alpha_2\alpha_3\phi_4$, $\alpha_2\phi_3\varphi$, $\alpha_3\phi_2\varphi$, and $\phi_2\phi_3\phi_4$, see Table~\ref{tab:frame_differences} for a schematic summary.
\begin{table}
\begin{center}
\begin{tabular}{c|c|c}
	& frame 1 & frame 2\\\hline
	$\alpha_2\alpha_3\phi_4$, $\phi_2\phi_3\phi_4$~ & ~$c_{ijk}\neq0$, $d_{ijk}\neq0$~ & ~$c_{ijk}=d_{ijk}=0$\\
	$\alpha_2\phi_3\varphi$, $\alpha_3\phi_2\varphi$~ & $c_{ijk}=0$, $d_{ijk}\neq0$ & ~$c_{ijk}=d_{ijk}=0$\\
\end{tabular}
\end{center}
\caption{Cubic couplings which are non-zero in frame 1 but vanish in frame 2.}
\label{tab:frame_differences}
\end{table}
On the other hand, we argued that the two frames are equivalent; after all, we just added a total derivative to the effective Lagrangian to move between the two frames. So how can it be that this does not lead to a contradiction?

There are two different answers to this question, depending on whether the cubic coupling is extremal or not:
\begin{itemize}
\item For $\phi_2\phi_3\phi_4$: this is a non-extremal coupling, and the apparent contradiction is resolved as follows. Even though both $c_{\phi_2\phi_3\phi_4}$ and $d_{\phi_2\phi_3\phi_4}$ (+ permutations) are non-zero in frame 1, their values are such that the \textit{total} contribution to the CFT 3pt-function coefficient happens to vanish due to a cancellation in the sum
\begin{align}\label{eq:Cp2p3p4}
	C_{\phi_2\phi_3\phi_4} =  c_{\phi_2\phi_3\phi_4}A_1+(d_{\phi_2\phi_3\phi_4}+d_{\phi_3\phi_2\phi_4}+d_{\phi_4\phi_2\phi_3})A_2=0\,.
\end{align}
To arrive at this result we used \eqref{eq:C_3pt} and \eqref{eq:norm_3pt}, together with the conformal dimensions summarized in Table~\ref{tab:PW_spectrum}.  On the other hand, in frame 2 the calculation is far simpler: both bulk cubic couplings are zero to start with. This then results in a trivially vanishing 3pt-function coefficient, in agreement with \eqref{eq:Cp2p3p4} computed in frame 1.
\item For $\alpha_2\alpha_3\phi_4$, $\alpha_2\phi_3\varphi$, $\alpha_3\phi_2\varphi$: these are all extremal couplings, for which the holographic computation is more subtle because of a divergence in the 3pt-Witten diagram. Indeed, recalling equation \eqref{eq:norm_3pt}, both $A_1$ and $A_2$ diverge for extremal arrangements of operator dimensions. There are different ways in which the divergence can be cancelled (see Section~\ref{subsec:extremal_couplings} for a more detailed discussion), but all cases essentially boil down to a situation of the type $0\cdot\infty$, and thus the corresponding CFT 3pt-function coefficients are inherently ambiguous. Considering a different frame of supergravity couplings then just corresponds to a different regularisation of the infinity. As discussed in Section~\ref{subsec:extremal_couplings} our proposal is that these extremal 3pt-functions vanish, provided that one carefully defines the notion of a single-particle operator in the dual CFT.
\end{itemize}

We now proceed to address yeet another subtlety in the calculation of holographic correlation functions, namely the contribution of finite terms near the asymptotic AdS boundary to the evaluation of Witten diagrams.

\subsubsection{Boundary terms}
\label{subsec:bdry}

So far we have discussed the cubic coupling that arise from the 5d bulk supergravity Lagrangian and how they can contribute to the evaluation of the 3pt-functions in the 4d boundary CFT. There are potential additional contributions to the CFT 3pt-functions arising from cubic terms that are finite at the boundary of AdS$_5$. While this possibility  is often overlooked when calculating correlation functions in AdS/CFT it is sometimes very important in order to find a precise agreement between the CFT and supergravity calculations, see \cite{Freedman:2016yue,mABJM} for a detailed discussion in the context of AdS$_4$ and the ABJM SCFT.

To address this question in the 5d supergravity model of interest in this work we need to study finite boundary terms built out of products of the 10 scalar fields. Since the AdS$_5$ backgrounds we study are supersymmetric, we also need to ensure that these finite boundary terms preserve supersymmetry. This is in general an involved technical problem to analyze rigorously in supergravity. There is however a well-known trick, often dubbed the Bogomol'nyi trick, which allows one to derive the result using a shortcut. The idea is that the finite boundary terms can be obtained by writing the supergravity action as a sum of squares plus a total derivative term that is built out of the superpotential of the model. This total derivative term then provides the finite boundary term of interest. While this is not a rigorous derivation of the finite boundary terms, experience suggests that this shortcut leads to correct results that have been tested in various models of precision holography and therefore we will also employ it in our analysis. The derivation of this Bogomol'nyi finite boundary term for the 10-scalar supergravity model of interest here was performed in \cite{Bobev:2016nua} and we will simply borrow these results. 

When the dust settles we find that both for the UV and IR AdS$_5$ solutions there are finite boundary terms built out of the scalar fields that could contribute to the evaluation of an on-shell action or holographic correlators. Concretely, these are polynomials in the scalar fields that scale as $e^{-4r/L}$ near the AdS boundary. For the UV AdS$_5$ backgrounds we find the following quadratic and cubic terms
\begin{align}\label{eq:bdy_terms_UV}
	\frac{1}{L}\bigg[3+\frac{1}{2}\big(\alpha_{1,2,3}^2+\beta_{1,2}^2\big)-\frac{1}{2\sqrt{6}}\big(\beta_1(\phi_1^2+\phi_2^2-2\phi_3^2)+\sqrt{3}\beta_2(\phi_1^2-\phi_2^2)\big)+(\text{quartic})\bigg]\,.
\end{align}
The quartic terms amount to 11 different combinations involving the fields $\phi_{1,2,3,4}$ which we do not present explicitly since they are not needed in our analysis. For the IR AdS$_5$ solution we find the following finite quadratic, cubic and quartic boundary terms that go as $e^{-4r/\LLS}$ near the asymptotic boundary
\begin{align}\label{eq:bdy_terms_IR}
	\frac{1}{\LLS}\bigg[3+\frac{1}{2}\bh^2+\frac{3}{4}\phi_2\phi_3\phi_4+\frac{9}{64}\phi_4^4\bigg]\,.
\end{align}

Given these finite boundary terms the next order of business is to understand how they contribute to the calculation of holographic 2pt- and 3pt-functions. As will be discussed in more detail in \cite{mABJM}, one can show that both the quadratic and cubic terms in \eqref{eq:bdy_terms_UV} and~\eqref{eq:bdy_terms_IR} do not contribute to correlation functions at separate points. They do however lead to $\delta$-functions contributions to the 2pt- and 3pt-functions, i.e. they lead to contact terms in the correlators. As we show in \cite{mABJM} such contact terms arise in holographic 2pt and 3pt-functions in AdS$_{d+1}$ from finite boundary terms when the conformal dimensions of the operators dual to the bulk scalar fields obey $\Delta_1 + \Delta_2 = d$ and $\Delta_1 + \Delta_2  + \Delta_3 = 2d$, respectively. Indeed, using \eqref{eq:bdy_terms_UV} and \eqref{eq:bdy_terms_IR}, the conformal dimensions in~\eqref{eq:masses_UV} and Table~\ref{tab:PW_spectrum} we find that the finite boundary terms above yield precisely such $\delta$-function contributions. Since our interest here is in correlation functions at separate points of the SCFTs on $\mathbb{R}^4$ we will not discuss these contact terms further. We nevertheless emphasize that they can play an important role in the calculation of integrated correlators of the theory on compact manifolds, for instance $S^4$.

\subsection{3pt-functions in the IR: predictions for the LS SCFT}
\label{subsec:3pt_results}
From the explicit results for the supergravity cubic couplings given above, the corresponding  3pt-function coefficients $C_{\Phi_i\Phi_j\Phi_k}$ are then easily obtained using the relation \eqref{eq:C_3pt}.\footnote{In \eqref{eq:C_3pt}-\eqref{eq:norm_3pt}, we used $\eta=\frac{N^2}{8\pi^2L^3}$ with $L=1$, recall \eqref{eq:S_eff}, and we reinstated the overall factor of $1/\LLS^2$ present in \eqref{eq:L_IR_exp}, with $\LLS$ given by $\LLS=\frac{3}{2^{5/3}}L$.}
Performing this calculation, one obtains a list of 31 distinct 3pt-functions. In order to organise our results, it is useful to group the various 3pt-functions according to the types of $\mathcal{N}=1$ multiplets their constituent operators belong to. Recall from Table~\ref{tab:PW_spectrum} that the 10 scalars are part of four different multiplets: two chiral multiplets ($\C$), the flavour current multiplet ($\F$), and a long multiplet ($\L$). Using these abbreviations, we have
\begin{align*}
	\{\alpha_2,\alpha_3,\phi_2,\phi_3\}\,,~\{\phi_4,\varphi\}\in\C\,,\quad \{\bh\}\in\F\,,\quad \{\rho_1,\alpha_1,\rho_2\}\in\L\,.
\end{align*}
The non-vanishing 3pt-functions then fall into one of the following 5 different classes: $\langle\C\C\C\rangle$, $\langle\C\C\F\rangle$, $\langle\C\C\L\rangle$, $\langle\F\F\L\rangle$, and $\langle\L\L\L\rangle$. We will label these classes with letters from (a) to (e). On the other hand, any other combination of multiplet types leads to a vanishing 3pt-function coefficient. We now list the explicit results, organised according to the above 5 different classes.
\begin{enumerate}[label=\bf{(\alph{enumi}})]
\item $\langle\C\C\C\rangle$: 
\begin{align}\label{eq:res_CCC}
	C_{\alpha_2\alpha_3\phi_4}=-\frac{2^{5/6}}{9}\,\frac{\gamma}{N}\,,\quad C_{\alpha_2\phi_3\varphi}=C_{\alpha_3\phi_2\varphi}=-\frac{1}{9\cdot2^{2/3}}\,\frac{\gamma}{N}\,,\quad C_{\phi_2\phi_3\phi_4}=0\,.
\end{align}
The first three 3pt-functions are extremal, and as discussed in Section \ref{subsec:extremal_couplings} their values are ambiguous and depend on the specific parametrisation of the bulk cubic couplings. We parametrise this ambiguity by $\gamma$,\footnote{In principle, one may even introduce different parameters $\gamma_i$ for each extremal 3pt-function coefficient, since their values can be adjusted \textit{independently} by means of \eqref{eq:Leff_shift}.} where $\gamma=1$ corresponds to the result obtained from the couplings in frame 1 (Section \ref{subsec:frame1}), whereas $\gamma=0$ corresponds to the computation being done in frame 2 (Section \ref{subsec:frame2}) where the extremal couplings vanish. Lastly, the 3pt-function coefficient $C_{\phi_2\phi_3\phi_4}$ is not extremal and vanishes independently of which frame is used, c.f. the discussion around equation \eqref{eq:Cp2p3p4}. All remaining 3pt-functions are also not extremal, leading to unambiguous results. As discussed in Section~\ref{subsec:extremal_couplings}, we propose that after properly defining single-particle operators in the CFT the extremal 3pt-functions vanish and thus $\gamma=0$. Nevertheless, we keep track of this potential ambiguity explicitly in our calculations.

\item $\langle\C\C\F\rangle$: 
\begin{align}\label{eq:res_CCF}
	C_{\alpha_2\alpha_2\bh}=-C_{\alpha_3\alpha_3\bh}=\frac{2\cdot2^{5/6}}{3}\,\frac{1}{N}\,,\qquad C_{\phi_2\phi_2\bh}=-C_{\phi_3\phi_3\bh}=-\frac{2\cdot2^{5/6}}{9}\,\frac{1}{N}\,.
\end{align}
%
\item $\langle\C\C\L\rangle$: the 3pt-functions with two operators from the $\Delta=\frac{3}{2}$ chiral multiplet read 
\begin{align}\label{eq:res_CCL1}
\begin{split}
	C_{\alpha_2\alpha_2\rho_1}&=C_{\alpha_3\alpha_3\rho_1}=C_{\alpha_2\phi_2\alpha_1}=C_{\alpha_3\phi_3\alpha_1}=\frac{\sqrt{1+\sqrt{7}}~\Gamma\big(\frac{2-\sqrt{7}}{2}\big)\Gamma\big(\frac{1+\sqrt{7}}{2}\big)}{2^{-\frac{10}{3}+\sqrt{7}}3^{\frac{3}{2}}\pi^{\frac{1}{2}}}\,\frac{1}{N}\,,\\
	C_{\phi_2\phi_2\rho_1}&=C_{\phi_3\phi_3\rho_1}=\frac{7-2\sqrt{7}}{3}\times C_{\alpha_2\alpha_2\rho_1}\,,\\
	C_{\alpha_2\alpha_2\rho_2}&=C_{\alpha_3\alpha_3\rho_2}=-\frac{\sqrt{7-2 \sqrt{7}}}{2\sqrt{14}}\times C_{\alpha_2\alpha_2\rho_1}\,,\\
	C_{\phi_2\phi_2\rho_2}&=C_{\phi_3\phi_3\rho_2}=-\frac{\sqrt{7+2\sqrt{7}}}{2\sqrt{6}}\times C_{\alpha_2\alpha_2\rho_1}\,.\\
\end{split}
\end{align}
When the two chiral operators belong to the $\Delta=3$ multiplet instead, we have
\begin{align}\label{eq:res_CCL2}
\begin{split}
	C_{\phi_4\phi_4\rho_1}&=-\frac{\sqrt{29+11\sqrt{7}}~\Gamma\big(\frac{5-\sqrt{7}}{2}\big)\Gamma\big(\frac{1+\sqrt{7}}{2}\big)^2\Gamma\big(\frac{3+\sqrt{7}}{2}\big)}{3\cdot 2^{\frac{1}{6}}\sqrt{21}\,\Gamma(\sqrt{7})}\,\frac{1}{N}\,,\\
	C_{\phi_4\phi_4\rho_2}&=-\frac{\sqrt{7-2 \sqrt{7}}}{2\sqrt{14}}\times C_{\phi_4\phi_4\rho_1}\,,\\
	C_{\varphi\varphi\rho_1}&=\frac{11-4 \sqrt{7}}{12}\times C_{\phi_4\phi_4\rho_1}\,,\\
	C_{\varphi\varphi\rho_2}&=-\frac{\sqrt{133+50\sqrt{7}}}{8\sqrt{42}}\times C_{\phi_4\phi_4\rho_1}\,,\\
	C_{\phi_4\varphi\alpha_1}&=\frac{1}{2\sqrt{2}}\times C_{\phi_4\phi_4\rho_1}\,.
\end{split}
\end{align}
Note that the 3pt-functions in the mixed case -- one operator from the $\Delta=\frac{3}{2}$, the other one from the $\Delta=3$ chiral multiplet -- vanish because of $\rm SU(2)_F$ selection rules.
\item $\langle\F\F\L\rangle$:
\begin{align}\label{eq:res_FFL}
\begin{split}
	C_{\bh\bh\rho_1}&=-\frac{2^{\frac{7}{3}}\sqrt{217+79\sqrt{7}}~\Gamma\big(\frac{3-\sqrt{7}}{2}\big)\Gamma\big(\frac{1+\sqrt{7}}{2}\big)^3}{9\sqrt{21\,\Gamma(\sqrt{7})\Gamma(1+\sqrt{7})}}\,\frac{1}{N},\\
	C_{\bh\bh\rho_2}&=\frac{\sqrt{91-34\sqrt{7}}}{2\sqrt{42}}\times C_{\bh\bh\rho_1}\,.
\end{split}
\end{align}
%
\item $\langle\L\L\L\rangle$:
\begin{align}\label{eq:res_LLL}
\begin{split}
	C_{\rho_1\rho_1\rho_1}&=\frac{2^{\frac{1}{3}}\sqrt{1659343-602999\sqrt{7}}~\Gamma\big(\frac{1+\sqrt{7}}{2}\big)^3\Gamma\big(\frac{-1+3\sqrt{7}}{2}\big)}{63 \sqrt{3}\big(\Gamma(\sqrt{7})\Gamma(1+\sqrt{7})\big)^{3/2}}\,\frac{1}{N}\,,\\
	C_{\rho_1\rho_1\rho_2}&=\frac{\sqrt{3(2777257+1042598\sqrt{7})}}{1514\sqrt{14}}\times C_{\rho_1\rho_1\rho_1}\,,\\
	C_{\rho_1\alpha_1\alpha_1}&=\frac{\sqrt{1976297+635680 \sqrt{7}}}{1514}\times C_{\rho_1\rho_1\rho_1}\,,\\
	C_{\rho_1\rho_2\rho_2}&=-\frac{4963+1156\sqrt{7}}{42392}\times C_{\rho_1\rho_1\rho_1}\,,\\
	C_{\rho_2\alpha_1\alpha_1}&=\frac{\sqrt{325648561+109746134\sqrt{7}}}{3028\sqrt{42}}\times C_{\rho_1\rho_1\rho_1}\,,\\
	C_{\rho_2\rho_2\rho_2}&=\frac{\sqrt{3(7587462463+2704517498\sqrt{7})}}{84784 \sqrt{2}}\times C_{\rho_1\rho_1\rho_1}\,.\\
\end{split}
\end{align}
\end{enumerate} 

A few comments are in order. Let us stress again that we do not know how to perform a direct computation of these structure constants using field theory methods. This is unlike the $\N=4$ SYM case, where 3pt-correlators of operators dual to supergravity fields obey non-renormalisation theorems and can be computed in the free theory limit (see also Appendix~\ref{app:3pt_UV}). The LS SCFT is different in mainly two regards. Firstly, there is no weakly coupled point on the conformal manifold where one could perform an analogous computation. Secondly, the supergravity spectrum consists not only of protected operators, but includes also unprotected, long ones. The above list of 3pt-function coefficients therefore constitutes a novel result, providing a valuable insight into the strongly coupled physics of the LS SCFT.

\section{Consistency checks on 3pt-functions}
\label{sec:3ptfnctsWard}

The aim of this section is to describe some consistency checks on the results presented in Section~\ref{subsec:3pt_results}. In particular, we show that the holographically computed 3pt-functions are consistent with various non-trivial constraints imposed by 4d $\N=1$ superconformal symmetry, and thus they agree with the expectations from the dual LS SCFT.

As a first check, one can show that all non-vanishing 3pt-functions are consistent with $\rm U(1)_R$ and $\rm SU(2)_F$ selection rules, see Table~\ref{tab:PW_spectrum}.\footnote{When checking the $\rm SU(2)_F$ selection rules we have used the decomposition of the following $\rm SU(2)$ tensor products $\mathbf{3}\otimes\mathbf{3}=(\mathbf{1}\oplus\mathbf{5})_S\oplus(\mathbf{3})_A$ and $\mathbf{3}\otimes\mathbf{3}\otimes\mathbf{3}=\mathbf{1}\oplus3\cdot\mathbf{3}\oplus2\cdot\mathbf{5}\oplus\mathbf{7}$.} In these considerations it is important to keep in mind that in our construction operators which are part of chiral multiplets $\C$ -- that is $\o_{\alpha_2}$, $\o_{\alpha_3}$, $\o_{\phi_4}$ and their descendants -- are linear combinations of the given multiplet and its complex conjugate , i.e. the chiral and anti-chiral multiplets. These operators therefore do not admit a definite $\rm U(1)_R$ R-charge assignment. This fact is a consequence of the particular 10-scalar model supergravity truncation we employ and somewhat limits our ability to fully exploit all constraints from $\N=1$ superconformal symmetry.

Bearing in mind this nuisance, we consider two types of consistency checks, both of which follow from leveraging the 4d $\N=1$ superconformal symmetry of the model. Generally speaking, the supersymmetry of the LS theory provides a relation between superconformal primaries $\o$ and their super-descendants. In the following, it will be useful to distinguish between level-2 and level-4 super-descendants, which we denote by $\o'$ and $\o''$, that can be obtained from $\o$ by the action of 2 or 4 supercharges, respectively. 
The two consistency checks we perform can then be schematically summarised as relations between
\begin{enumerate}[label=($\roman{enumi}$)]
\item $\langle\o\o'\o'\rangle$ and $\langle\o\o\o\rangle$ from computing norms of level-2 descendants;
\item $\langle\o\o\o''\rangle$ and $\langle\o\o\o\rangle$ from exploiting superconformal blocks.
\end{enumerate}
The details of these consistency checks are discussed in Sections~\ref{subsec:check1} and \ref{subsec:check2} below.

\subsection{Relating $\langle\o\o'\o'\rangle$ to $\langle\o\o\o\rangle$ via level-2 norms}
\label{subsec:check1}

The basic idea is to use the operator-state correspondence valid for every Euclidean CFT to evaluate the 3pt-function $\langle\o_1'\vert\o\vert\o_2'\rangle=\langle\o_1\vert S S \o Q Q\vert\o_2\rangle$. On the right hand side we have used that the conjugate of a $Q$ Poincar\'e supercharge is an $S$ conformal supercharge, see Section 2.1 in \cite{Bobev:2015jxa} for more details on our conventions. We can then use the 4d $\mathcal{N}=1$ superconformal algebra to  move the $S$ supercharges to the right and relate this 3pt-function to the 3pt-function of  the primary operators $\langle\o_1\vert\o\vert\o_2\rangle$. Importantly, when performing this calculation we should keep in mind the radial ordering of operators dictated by the operator-state correspondence, we take $|y|>|x|>0$. In addition, we should impose that the operator in the middle, $\o$, is a chiral (or anti-chiral) superconformal primary $\mathcal{C}$, i.e. it is annihilated by half of the $Q$ supercharges, and the operator on the right, $\o_2$, belongs to a short multiplet, i.e. is annihilated by the simultaneous action of all four $Q$ supercharges. This analysis leads to the explicit calculation of the ratio
\begin{align}\label{eq:level2norms}
	\frac{\langle\o_1(y)\vert \mathcal{C}(x)\vert\o_2(0)\rangle}{\langle Q^-Q^-\o_1(y)\vert \mathcal{C}(x)\vert Q^+Q^+\o_2(0)\rangle}\,,
\end{align}
derived using the superconformal algebra and the fact that  $\o_1$ and $\o_2$ are \textit{scalar} superconformal primaries. When performing this calculation we should keep in mind that in our model the operators dual to the supergravity scalar fields are linear combinations of chiral and anti-chiral primaries. We also need to use the relation $\mathcal{C}(x) = {\rm e}^{{\rm i}P.x}\mathcal{C}(0){\rm e}^{-{\rm i}P.x}$ and the following norm of the level-2 descendants, see Equation (36) of \cite{Bobev:2015jxa} specialised to $d=4$, which can be obtained using the superconformal algebra
\begin{align}\label{eq:normlev2}
	\langle\o'_{\pm}\vert\o'_{\pm}\rangle = 4\big(\Delta\mp\tfrac{3}{2}q\big)\big(\Delta\mp\tfrac{3}{2}q-2\big)\,.
\end{align} 
Here $\vert\o\rangle$ is a superconformal scalar primary of dimension $\Delta$ and $\rm U(1)_R$ R-charge $q$ and $\vert\o'_{\pm}\rangle=\epsilon^{\alpha\beta}Q_\alpha^{\pm}Q_\beta^{\pm}\vert\o\rangle$ is the level-2 descendant. 

\paragraph{Checks on $\langle\C\C\L\rangle$ 3pt-functions:}
To illustrate the calculation let us take the operator $\mathcal{C}$ in \eqref{eq:level2norms} to be the one dual to the scalar field $\phi_4$, the operator $\o_1$ to be the dual to $\rho_1$ and the operator $\o_2$ to be again the dual to $\phi_4$. The corresponding superconformal primary states are
\begin{equation}
|\rho_1\rangle = |0,0\rangle\,, \qquad |\phi_4\rangle = \frac{1}{\sqrt{2}}\left(|0,2\rangle+|0,-2\rangle\right)\,.
\end{equation}
Here we have used the notation $|m,n\rangle$ to denote a unit norm state with $\rm SU(2)_F$ weight $m$ and $\rm U(1)_R$ charge $n$. We thus have that both $|\rho_1\rangle$ and $|\phi_4\rangle$ are states with unit norm. The level two descendants of these states then take the form 
\begin{equation}
\begin{split}
|\alpha_1\rangle &= \xi_1 \left(\epsilon^{\dot{\alpha}\dot{\beta}}Q^{-}_{\dot{\alpha}}Q^{-}_{\dot{\beta}}|0,0\rangle + \epsilon^{\alpha\beta}Q^{+}_{\alpha}Q^{+}_{\beta}|0,0\rangle\right)\,, \\
 |\varphi\rangle &= \xi_2 \left(\epsilon^{\dot{\alpha}\dot{\beta}}Q^{-}_{\dot{\alpha}}Q^{-}_{\dot{\beta}}|0,2\rangle + \epsilon^{\alpha\beta}Q^{+}_{\alpha}Q^{+}_{\beta}|0,-2\rangle\right)\,,
 \end{split}
\end{equation}
where $\xi_{1,2}$ are normalization constants. To evaluate the norms of these descendant states we can use \eqref{eq:normlev2} and the conformal dimensions from Table~\ref{tab:PW_spectrum} to find
\begin{equation}
\langle\alpha_1|\alpha_1\rangle = 8 \xi_1^2\Delta_{\rho_1}(\Delta_{\rho_1}-2)=48\xi_1^2\,, \qquad  \langle\varphi|\varphi\rangle = 64 \xi_2^2\Delta_{\phi_4}(\Delta_{\phi_4}-2)=192\xi_2^2\,.
\end{equation}
In order to have unit normalized 2pt-functions we need to fix the constants $\xi_{1,2}$ such that the states $|\alpha_1\rangle$ and $|\varphi\rangle$ have unit norm, i.e. 
\begin{equation}
\xi_1=\frac{1}{\sqrt{48}}\,, \qquad \xi_2=\frac{1}{\sqrt{192}}\,.
\end{equation}
Using the 4d $\mathcal{N}=1$ superconformal algebra and the definitions and normalization above we then find the following ratio of 3pt-functions
\begin{equation}
\frac{\langle\rho_1|\phi_4|\phi_4\rangle}{\langle\alpha_1|\phi_4|\varphi\rangle} = \frac{1}{24\sqrt{2}\,\xi_1\xi_2} = 2\sqrt{2}\,.
\end{equation}
This result agrees nicely with the holographic calculation of the corresponding correlators where from \eqref{eq:res_CCL2} we find
\begin{align}
	\frac{C_{\rho_1\phi_4\phi_4}}{C_{\alpha_1\phi_4\varphi}}=2\sqrt{2}\,.
\end{align}
Performing an analogous calculation one finds the following ratios of correlation functions
\begin{equation}
\frac{\langle\rho_1|\alpha_2|\alpha_2\rangle}{\langle\alpha_1|\alpha_2|\phi_2\rangle} = \frac{\langle\rho_1|\alpha_3|\alpha_3\rangle}{\langle\alpha_1|\alpha_3|\phi_3\rangle}=1\,,
\end{equation}
which also agrees with the holographic results in \eqref{eq:res_CCL1}.
 
\paragraph{Checks on $\langle\C\C\C\rangle$ 3pt-functions:}
We can perform one more consistency check on our calculations employing the method described above. To this end we can take all three superconformal primaries to be chiral/anti-chiral operators and calculate the ratios
\begin{equation}\label{eq:extrCCCcheck}
\frac{\langle\alpha_3|\alpha_2|\phi_4\rangle}{\langle\phi_3|\alpha_2|\varphi\rangle} =\frac{\langle\alpha_2|\alpha_3|\phi_4\rangle}{\langle\phi_2|\alpha_3|\varphi\rangle} =2\sqrt{2}\,.
\end{equation}
This result agrees with the holographic calculation in \eqref{eq:res_CCC} for any finite value of the parameter $\gamma$. As discussed above, the single-particle operator prescription determines the value $\gamma=0$, but it is nevertheless encouraging that for any value of $\gamma$ the superconformal Ward identity that determines the ratio in \eqref{eq:extrCCCcheck} is obeyed.

If we instead take $\phi_4$ to be the operator in the middle, we find that the 3pt-function of level-2 descendants vanishes:
\begin{align}
	\langle\phi_2|\phi_4|\phi_3\rangle = 0\,,
\end{align}
which is in agreement with the holographic computation quoted in the last equation in~\eqref{eq:res_CCC}. In fact, as discussed in Section~\ref{subsec:bdry} above, this is only true at separated points, i.e. up to contact terms, due to finite cubic boundary term of the form $\int d^4x\, \phi_2\phi_3\phi_4$, see \eqref{eq:bdy_terms_IR}.

\subsection{Relating $\langle\o\o\o''\rangle$ to $\langle\o\o\o\rangle$ via superconformal blocks}
\label{subsec:check2}

The 4d $\N=1$ superconformal blocks, denoted here schematically by $\mathcal{G}_{\Delta,\ell}$, are supersymmetric extensions of the usual conformal blocks $g_{\Delta,\ell}$ which, in addition to the usual conformal descendants, also take into account contributions from all \textit{super}-descendants of the superconformal primary with dimension $\Delta$ and spin $\ell$ to a given 4pt-function. Therefore, superconformal blocks encode information about certain ratios of 3pt-functions involving primaries and a super-descendants which we can exploit to perform consistency checks on our holographic calculations.

For a generic 4pt-function of scalar primary operators $\langle\o_1\o_2\o_3\o_4\rangle$ in a 4d $\N=1$ SCFT, the superconformal block associated to the exchange of a long supermultiplet takes the schematic form
\begin{align}\label{eq:superblock}
	\mathcal{G}_{\Delta,\ell} = g_{\Delta,\ell}+c_1\,g_{\Delta+1,\ell+1}+c_2\,g_{\Delta+1,\ell-1}+c_3\,g_{\Delta+2,\ell}\,.
\end{align}
The $c_{i}$ are coefficients which depend on the dimension $\Delta$ and spin $\ell$ of the exchanged operator, as well as the type of multiplet the external operators $\o_i$ belong to, i.e. $\o_i\subset\mathcal{M}_i$ with $\mathcal{M}_i\in\{\C,\F,\L\}$. Here we will be interested in the superconformal blocks for 4pt-functions of chiral multiplets $\langle\C\C\C\C\rangle$, flavor multiplets $\langle\F\F\F\F\rangle$, and long multiplets $\langle\C\C\L\L\rangle$ and $\langle\L\L\L\L\rangle$. For these cases, the relevant superconformal blocks have been derived in the literature and can be found in e.g. \cite{Li:2017ddj}. Moreover, we will only focus on the coefficient $c_3$ in \eqref{eq:superblock}, which encodes the relation between the 3pt-function of the exchanged superconformal primary operator and its level-4 supersymmetric descendant. We employ the following notation to denote this coefficient 
\begin{align}\label{eq:c3_relation}
	c_3 \equiv c_{\Delta,\ell}^{(\mathcal{M}_1\mathcal{M}_2\mathcal{M}_3\mathcal{M}_4)} = \frac{C_{\o_1\o_2\o''}\,C_{\o''\o_3\o_4}}{C_{\o_1\o_2\o}\,C_{\o\o_3\o_4}}\,,
\end{align}
where $\o$ is an exchanged unprotected superconformal primary operator with quantum numbers $(\Delta,\ell)$, and $\o''\sim Q^2\bar{Q}^2\o$ denotes its level-4 descendant. Since there is only one long primary operator in the spectrum of the 10-scalar model we study here, the role of $\o$ in our setup is played by $\o_{\rho_1}$ with $(\Delta,\ell)=(1+\sqrt{7},0)$, and $\o''$ is the level-4 descendant $\o_{\rho_2}$. Using the explicit form of the 4d $\N=1$ superconformal blocks  we can then find the coefficients $c_{\Delta,\ell}^{(\mathcal{M}_1\mathcal{M}_2\mathcal{M}_3\mathcal{M}_4)}$ for the various types of 4pt-functions mentioned above, and check that the relation in \eqref{eq:c3_relation} is satisfied by the holographic 3pt-functions presented in Section~\ref{subsec:3pt_results}.

\paragraph{Checks on $\langle\C\C\L\rangle$ 3pt-functions:} The expression for the $\langle\bar{\C}\C\bar{\C}\C\rangle$ superconformal block can be found in (2.7)-(2.8) of \cite{Li:2017ddj}.\footnote{In our setup the external operators are an admixture of chiral $\mathcal{C}$ and anti-chiral $\bar{\C}$ operators. Nevertheless, it is possible to use the superconformal blocks in (2.7)-(2.8) of \cite{Li:2017ddj} because chiral-chiral-descendant 3pt-functions vanish for level-4 descendants.} Using this we can read off the coefficient $c_3$
\begin{equation}
c_3 = \frac{1}{16} \frac{(\Delta+\ell)(\Delta-\ell-2)}{(\Delta^2-(\ell+1)^2)}\,.
\end{equation}
Taking the external operators to be all $\o_{\alpha_2}$ and the superconformal primary operator in the OPE to be $\rho_1$ we find
\begin{align}\label{eq:check_CCL}
	c_{\Delta=1+\sqrt{7},\ell=0}^{(\bar{\C}\C\bar{\C}\C)} = \frac{7-2 \sqrt{7}}{56} \overset{!}{=} \left(\frac{C_{\alpha_2\alpha_2\rho_2}}{C_{\alpha_2\alpha_2\rho_1}}\right)^2\,,
\end{align}
where on the right hand side we have indicated that the ratio agrees with the holographic result in \eqref{eq:res_CCL1}. The same consistency check applies to the case when the external operators are $\o_{\alpha_3}$ or $\o_{\phi_4}$. Indeed, from the results in \eqref{eq:res_CCL1} and \eqref{eq:res_CCL2} we have
\begin{align}\label{eq:chiral_ratios}
	\frac{C_{\alpha_2\alpha_2\rho_2}}{C_{\alpha_2\alpha_2\rho_1}}=\frac{C_{\alpha_3\alpha_3\rho_2}}{C_{\alpha_3\alpha_3\rho_1}}=\frac{C_{\phi_4\phi_4\rho_2}}{C_{\phi_4\phi_4\rho_1}}\,.
\end{align}
%

\paragraph{Checks on $\langle\F\F\L\rangle$ 3pt-functions:} The superconformal block for external scalars belonging to a flavor current multiplet, i.e. for the 4pt-function $\langle\F\F\F\F\rangle$, can be found in (3.7) of \cite{Li:2017ddj} and leads to
\begin{equation}
c_3 = \frac{1}{16} \frac{(\Delta-2)^2(\Delta+\ell)(\Delta-\ell-2)}{\Delta^2(\Delta^2-(\ell+1)^2)}\,.
\end{equation}
Taking the external operators to be $\o_{\bh}$ and the propagating superconformal primary to be $\o_{\rho_1}$, we obtain\footnote{We note there is a mutual consistency of the $\langle\bar{\C}\C\bar{\C}\C\rangle$ and $\langle\F\F\F\F\rangle$ superconformal blocks with the $\langle\bar{\C}\C\F\F\rangle$ superconformal block in (3.4) of \cite{Li:2017ddj} due to the relation $c_{\Delta,\ell}^{(\bar{\C}\C\F\F)}=\sqrt{c_{\Delta,\ell}^{(\bar{\C}\C\bar{\C}\C)}c_{\Delta,\ell}^{(\F\F\F\F)}}$.}
\begin{align}
	c_{\Delta=1+\sqrt{7},\ell=0}^{(\F\F\F\F)} = \frac{91-34 \sqrt{7}}{168} \overset{!}{=} \left(\frac{C_{\bh\bh\rho_2}}{C_{\bh\bh\rho_1}}\right)^2.
\end{align}
This again nicely agrees with our holographic result in \eqref{eq:res_FFL}.

\paragraph{Checks on $\langle\L\L\L\rangle$ 3pt-functions:} A slightly more intricate consistency check can be performed for the 3pt-function involving three long multiplets. It turns out that the $\langle\L\L\L\L\rangle$ superconformal block is not fully determined by $\N=1$ superconformal symmetry. Instead, the coefficient $c_3$ is fixed only up to an unknown ratio of 3pt-functions, see (2.41) of \cite{Li:2017ddj}, and reads
\begin{equation}\label{eq:c3LLLL}
c_{3}^{(\L\L\L\L)}=\frac{[(\Delta+\ell)^2 - 8 (\Delta-1)\frac{c_{LLL}^{(2)}}{c_{LLL}}]^2}{16\Delta^2(\Delta-\ell-1)(\Delta-\ell-2)(\Delta+\ell)(\Delta+\ell+1)}\,.
\end{equation}
However, the same undetermined ratio appears in the coefficient $c_3$ of the $\langle\bar{\C}\C\L\L\rangle$ superconformal block in (2.38) of \cite{Li:2017ddj}
\begin{equation}\label{eq:c3CCLL}
c_{3}^{^{(\bar{\C}\C\L\L)}}=\frac{(\Delta+\ell)^2 - 8 (\Delta-1)\frac{c_{LLL}^{(2)}}{c_{LLL}}}{16\Delta(\Delta^2-(\ell+1)^2)}\,.
\end{equation}
We can exploit this and first take the $\langle\bar{\C}\C\L\L\rangle$ 4pt-function for the operators $\o_{\alpha_2}$ and $\o_{\rho_1}$\footnote{Due to the equality \eqref{eq:chiral_ratios} we can equivalently use $\o_{\alpha_3}$ or $\o_{\phi_4}$ as the external chiral primaries.} to  find
\begin{align}
	c_{\Delta=1+\sqrt{7},\ell=0}^{(\bar{\C}\C\L\L)} = \frac{C_{\alpha_2\alpha_2\rho_2}C_{\rho_2\rho_1\rho_1}}{C_{\alpha_2\alpha_2\rho_1}C_{\rho_1\rho_1\rho_1}}\,.
\end{align}
Plugging in the value of $\frac{C_{\alpha_2\alpha_2\rho_2}}{C_{\alpha_2\alpha_2\rho_1}}$, which we already checked in \eqref{eq:check_CCL}, we can find the undetermined ratio $\frac{c_{LLL}^{(2)}}{c_{LLL}}$ in \eqref{eq:c3CCLL} and plug it back into the coefficient \eqref{eq:c3LLLL} of the $\langle\L\L\L\L\rangle$ superconformal block. This leads to the following result
\begin{align}
	c_{\Delta=1+\sqrt{7},\ell=0}^{(\L\L\L\L)} = \frac{3 \left(2777257+1042598 \sqrt{7}\right)}{32090744} \overset{!}{=} \left(\frac{C_{\rho_1\rho_1\rho_2}}{C_{\rho_1\rho_1\rho_1}}\right)^2\,.
\end{align}
Yet again, this agrees with the holographic 3pt-functions presented in \eqref{eq:res_LLL}.

\subsection{Comments on other 3pt-functions}

There are some 3pt-functions of operators in our 10-scalar consistent truncation which are allowed by the $\rm U(1)_R$ and $\rm SU(2)_F$ selection rules but vanish. Here we discuss these correlators.

Some of the operators dual to the fields in the 10-scalar supergravity model are odd under parity, i.e. they are pseudo-scalars. Since the LS SCFT is parity invariant, correlation functions with an odd number of pseudo-scalar operators must vanish. To understand which of the scalars in our model are odd under parity we note that any superconformal descendant obtained by acting with only $Q$ or only $\bar{Q}$ supercharges on a superconformal primary invariant under parity is parity odd. In the 10-scalar models the operators dual to the scalar fields $\varphi$, $\phi_{2,3}$, and $\alpha_1$ are precisely such parity-odd superdescendants. The other six scalars in the 10-scalar model are dual to parity-even operators. We therefore conclude that the parity invariance of the LS SCFT implies that the following 3pt-functions must vanish 
\begin{itemize}
\item 3 pseudo-scalars: $C_{\alpha_1\alpha_1\varphi}$, $C_{\alpha_1\phi_{2,3}\phi_{2,3}}$, $C_{\varphi\varphi\varphi}$, $C_{\varphi\phi_{2,3}\phi_{2,3}}$.
\item 1 pseudo-scalar: $C_{\bh\bh\varphi}$, $C_{\bh\alpha_{2,3}\phi_{2,3}}$, $C_{\rho_{1,2}\alpha_{2,3}\phi_{2,3}}$, $C_{\rho_{1,2}\rho_{1,2}\varphi}$, $C_{\rho_{1,2}\alpha_1\phi_4}$, $C_{\alpha_1\alpha_{2,3}\alpha_{2,3}}$, $C_{\varphi\phi_4\phi_4}$,

\vspace{-0.1cm}\hspace{3.05cm}$C_{\varphi\alpha_{2,3}\alpha_{2,3}}$, $C_{\phi_4\alpha_{2,3}\phi_{2,3}}$.
\end{itemize}
Indeed, we find that our supergravity effective action leads to the vanishing of these 3pt-functions.

These two sets of vanishing 3pt-functions, however, do not exhaust the full list. We in addition find that the following 3pt-functions are allowed by the parity and $\rm U(1)_R\times\rm SU(2)_F$ selection rules but happen to vanish in the 10-scalar model, i.e. there are no corresponding cubic terms in \eqref{eq:P3_IR_new}.
\begin{itemize}
\item Vanishing cubic couplings:\vspace{0.1cm}\\\vspace{0.1cm}
- $\langle\F\F\F\rangle$: $C_{\bh\bh\bh}$,\\\vspace{0.1cm}
- $\langle\C\C\F\rangle$: \hspace{0.2cm}$C_{\alpha_2\alpha_3\bh}$, $C_{\phi_2\phi_3\bh}$,\\\vspace{0.1cm}
- $\langle\C\C\L\rangle$: \hspace{0.25cm}$C_{\alpha_2\alpha_3\rho_{1,2}}$,\\\vspace{0.1cm}
- $\langle\C\C\C\rangle$: \hspace{0.3cm}$C_{\alpha_1\alpha_2\phi_3}$, $C_{\alpha_1\alpha_3\phi_2}$, $C_{\alpha_2\phi_2\varphi}$,$C_{\alpha_3\phi_3\varphi}$, $C_{\alpha_2\alpha_2\phi_4}$, $C_{\alpha_3\alpha_3\phi_4}$, $C_{\phi_2\phi_2\phi_4}$, $C_{\phi_3\phi_3\phi_4}$.
\end{itemize}

It will be very interesting to understand the reason for the vanishing of these 3pt-functions. One natural guess is that there may be a selection rule that we have overlooked, similar to the $\rm U(1)_Y$ ``bonus symmetry'' that explains some of the vanishing correlation functions in the large-$N$ limit of the $\mathcal{N}=4$ SYM theory \cite{Intriligator:1998ig}. Alternatively, these correlation functions may vanish only to leading order in the $1/N$ expansion and acquire non-zero values at finite $N$.

\section{Another truncation and higher-point correlators}
\label{sec:higher-pt}

The 10-scalar supergravity truncation studied above is a simple and useful model which allowed us to compute 3pt-functions of scalar operators. In this section we comment on the holographic computation of higher-point functions for the LS SCFT. To this end, we introduce another supergravity truncation which besides scalars also contains gauge fields. This is necessary in order to compute higher-point functions, as vector fields can in principle be exchanged in 4pt-functions of external scalars. Moreover, using this other truncation we compute some of the already obtained 3pt-functions of scalars in an independent fashion, thus providing a further consistency check of our results.

\subsection{A supergravity truncation with gauge fields}
\label{subsec:another_truncation}

The model we will consider is a consistent truncation of the 5d $\mathcal{N}=8$ maximal gauged supergravity obtained by imposing invariance under certain discrete symmetries of the theory, see \cite{Bobev:2010de} for the detailed derivation of this model. In addition to the 5d metric, the bosonic fields in the model are 3 ${\rm U}(1)$ gauge fields, $\Am^{(1,2,3)}$, together with 10 scalar fields, $\beta_{1, 2}\,$, $\phi_{1,2,3,4}$, and $\theta_{1,2,3,4}\,$. The metric, the gauge fields and the $\beta_{1,2}$ scalars organize into the bosonic fields of the well-known STU model of gauged supergravity, see for example \cite{Cvetic:1999xp}. The other 8 scalar fields arise as a subsector of 4 hypermultiplets coupled to this STU model.\footnote{Note that compared to \cite{Bobev:2010de} we renamed some of the fields in order to match the notation of the 10-scalar model introduced in Section \ref{sec:10-scalar}. In particular, we replaced $\alpha\mapsto-\beta_1$, $\beta\mapsto-\beta_2$ and $\varphi_j\mapsto\phi_j$.} The Lagrangian $\Lt$ of this supergravity truncation reads \cite{Bobev:2010de}:
\begin{align}\label{eq:Lagrangian_2}
\begin{split}
	\Lt &= -R-\Big[\rho^4\nu^{-4}\,\Fmn^{(1)}\Fmnup{1}+\rho^4\nu^4\,\Fmn^{(2)}\Fmnup{2}+\rho^{-8}\,\Fmn^{(3)}\Fmnup{3}\Big]\\
	&\quad+12(\parm\beta_1)^2+4(\parm\beta_2)^2+2\sum_{j=1}^4(\parm\phi_j)^2\\
	&\quad+\frac{1}{2}\sinh^2(2\phi_1)\Big(\parm\theta_1+(\Am^{(1)}+\Am^{(2)}-\Am^{(3)})\Big)^2\\
	&\quad+\frac{1}{2}\sinh^2(2\phi_2)\Big(\parm\theta_1+(\Am^{(1)}-\Am^{(2)}+\Am^{(3)})\Big)^2\\
	&\quad+\frac{1}{2}\sinh^2(2\phi_3)\Big(\parm\theta_1+(-\Am^{(1)}+\Am^{(2)}+\Am^{(3)})\Big)^2\\
	&\quad+\frac{1}{2}\sinh^2(2\phi_4)\Big(\parm\theta_1-(\Am^{(1)}+\Am^{(2)}+\Am^{(3)})\Big)^2+\Pt\,,\\
\end{split}
\end{align}
where we introduced $\rho= e^{-\beta_1}$ and $\nu= e^{-\beta_2}$, and $\Fmn^{(J)}$ denote the field strengths of the gauge fields $\Am^{(J)}$. The scalar potential $\Pt$ can be compactly written as
\begin{align}\label{eq:potential_2}
	\Pt=\frac{1}{2L^2}\bigg[\,\sum_{j=1}^4 \left(\frac{\partial W}{\partial\phi_j}\right)^2+\frac 16\left(\frac{\partial W}{\partial\beta_1}\right)^2+\frac 12\left(\frac{\partial W}{\partial\beta_2}\right)^2\,\bigg]-\frac{4}{3L^2}W^2\,,
\end{align}
with the ``superpotential'' $W$ given by
\begin{align}
\begin{split}
	W=&-\frac{1}{2\rho^2\nu^2}\bigg[(1+\nu^4-\nu^2\rho^6)\cosh(2\phi_1)+(-1+\nu^4+\nu^2\rho^6)\cosh(2\phi_2)\\
	&\qquad\quad\,+(1-\nu^4+\nu^2\rho^6)\cosh(2\phi_3)+(1+\nu^4+\nu^2\rho^6)\cosh(2\phi_4)\bigg].
\end{split}
\end{align}

There are again two supersymmetric critical points of the potential, whose holographic duals are the $\mathcal{N}=4$ SYM theory and the $\mathcal{N}=1$ LS SCFT, respectively. For both critical points, we now summarize the spectrum of linearized supergravity fluctuations, as well as the corresponding dual CFT operators and their 3pt-functions. The computations that lead to these results are analogous to what we presented in Sections~\ref{sec:10-scalar} and~\ref{sec:3pt_IR}, so we will keep the discussion brief.

\paragraph{The UV vacuum:}
This critical point is located at the origin of the scalar manifold. The six scalars and the potential take the values
\begin{align}\label{eq:sol_UV_2}
	\beta_{1,2}=\phi_{1,2,3,4}=0\,,\quad\text{with}~~\Pt=-\frac{12}{L^2}\,.
\end{align}
Expanding the above effective Lagrangian to quadratic order, one finds the following mass spectrum for the scalars:
\begin{equation}\label{eq:masses_UV2}
\begin{alignedat}{2}
	\beta_{1,2}:&\quad m^2L^2=-4\qquad  &\leftrightarrow\qquad \Delta&=2\,,\\
	\phi_{1,2,3,4}:&\quad m^2L^2=-3 &\leftrightarrow\qquad \Delta&=3\,,
\end{alignedat}
\end{equation}
in agreement with a subset of the mass spectrum found in the 10-scalar model, c.f. \eqref{eq:masses_UV}. The four scalars $\theta_{1,2,3,4}$ are massless and together with the scalars $\phi_{1,2,3,4}$ can be organized into complex scalars that are dual to four complex fermionic bilinear operators in the $\mathbf{10}\oplus\overline{\mathbf{10}}$ of the ${\rm SU}(4)$ R-symmetry of the $\mathcal{N}=4$ SYM theory. The new ingredients in this truncation, the three gauge fields $\Am^{(J)}$, are massless in the UV, i.e.
\begin{equation}
\begin{alignedat}{2}
	\Am^{(1,2,3)}:&\quad m^2L^2=0\qquad  &\leftrightarrow\qquad \Delta&=3\,,
\end{alignedat}
\end{equation}
and correspond to spin-1 operators of dimension $\Delta=3$ in $\N=4$ SYM theory. These are simply the three conserved currents corresponding to the Cartan subalgebra of the ${\rm SU}(4)$ R-symmetry.

To compute the 3pt-functions we need to expand the Lagrangian in~\eqref{eq:Lagrangian_2} to cubic order in the small fluctuations. We find only two non-zero scalar cubic couplings of the form $\beta_1^3$ and $\beta_1\beta_2^2$. These give rise to the following 3pt-function coefficients
\begin{align}
	C_{\beta_1\beta_1\beta_1}=-C_{\beta_1\beta_2\beta_2}=-\sqrt{\frac{2}{3}}\,\frac{1}{N}\,,
\end{align}
which matches the results from the 10-scalar model presented in Appendix \ref{app:3pt_UV}, c.f. eq. \eqref{eq:3pt_UV}.

\paragraph{The IR vacuum:}
At the second supersymmetric critical point of the potential, some of the scalars take non-zero expectation values given by
\begin{align}\label{eq:sol_IR_2}
	\beta_2=\phi_{2,3,4}=0\,,\quad\beta_1=-\frac{\log 2}{6}\,,\quad\phi_1=\frac{\log3}{2}\,,\quad\text{with}~~\Pt=-\frac{12}{\LLS^2}\,,
\end{align}
where $\LLS$ is defined in \eqref{eq:sol_IR}. Expanding the Lagrangian \eqref{eq:Lagrangian_2} around this solution to quadratic order and diagonalising the mass matrix by an appropriate change of basis $\{\beta_1,\phi_1\}\mapsto\{\rho_1,\rho_2\}$ along the lines of Section \ref{subsec:spectrum_IR} (and renaming $\beta_2\mapsto\bh$), leads to the following spectrum of scalar fluctuations: 
\begin{equation}\label{eq:spectrum_scalar2}
\begin{alignedat}{2}
	\bh:&\quad m^2\LLS^2=-4 &\leftrightarrow\qquad \Delta&=2\,,\\
	\phi_{2,3}:&\quad m^2\LLS^2=-\tfrac{15}{4} &\leftrightarrow\qquad \Delta&=\tfrac{5}{2}\,,\\
	\phi_{4}:&\quad m^2\LLS^2=-3 &\leftrightarrow\qquad \Delta&=3\,,\\[-3pt]
	\rho_1:&\quad m^2\LLS^2=4-2\sqrt{7}\qquad  &\leftrightarrow\qquad \Delta&=1+\sqrt{7}\,,\\[-3pt]
	\rho_2:&\quad m^2\LLS^2=4+2\sqrt{7} &\leftrightarrow\qquad \Delta&=3+\sqrt{7}\,.
\end{alignedat}
\end{equation}
Again, this agrees with our previous findings for 10-scalar model summarised in Table~\ref{tab:PW_spectrum}, apart from the fields $\{\alpha_{1,2,3},\varphi\}$ which are absent in this truncation. Similarly, one finds a non-diagonal mass matrix also for the gauge fields. Performing the necessary change of basis $\Am^{(J)}\mapsto\widetilde{A}_\mu^{(J)}$ we find the spin-1 spectrum to be given by
\begin{equation}
\begin{alignedat}{2}
	\widetilde{A}_\mu^{(1,2)}:&\quad m^2\LLS^2=0\qquad  &\leftrightarrow\qquad \Delta&=3\,,\\
	\widetilde{A}_\mu^{(3)}:&\quad m^2\LLS^2=6\qquad  &\leftrightarrow\qquad \Delta&=2+\sqrt{7}\,,\\
\end{alignedat}
\end{equation}
i.e. two vector fields remain massless, while $\widetilde{A}_\mu^{(3)}$ acquires a non-zero mass. The massless gauge fields are dual to the Cartan generators of the ${\rm SU}(2)_{\rm F}\times {\rm U}(1)_{\rm R}$ global symmetry of the LS SCFT.\footnote{See \cite{Bobev:2014jva} for yet another 5d supergravity truncation that retains all gauge fields dual to the ${\rm SU}(2)_{\rm F}\times {\rm U}(1)_{\rm R}$ global symmetry of the LS theory.} The massive spin-1 operator dual to $\widetilde{A}_\mu^{(3)}$ belongs to the long multiplet $L\bar{L}[1+\sqrt{7};0,0;0]\otimes[0]^{(0)}$, see Table~\ref{tab:PW_spectrum}, of which $\o_{\rho_1}$ is the superprimary. The scalars $\theta_{2,3,4}$ can be thought of as complex phases that can be combined with $\phi_{2,3,4}$ into appropriate complex operators in the $L\overline{B}_1$ multiplets in Table~\ref{tab:PW_spectrum}. The scalar $\theta_1$ is ``eaten up'' by the $\widetilde{A}_\mu^{(3)}$ massive gauge field via the Higgs mechanism.

Performing the expansion of the scalar modes to cubic order we find, in contrast to the previous 10-scalar model, that there are no cubic or higher-order contributions from the kinetic terms of the scalars, and hence all derivative couplings $d_{ijk}$ are manifestly absent. In the nomenclature of Section~\ref{subsec:frame2}, we are thus directly in the analogue of `frame 2' for the supergravity couplings and we again observe the vanishing of all extremal couplings. Furthermore, we find that the numerical coefficients of all cubic couplings in this model agree with the ones of the 10-scalar model given in~\eqref{eq:P3_IR_new}, after restricting to the subset of scalars~\eqref{eq:spectrum_scalar2}. This implies that also the dual 3pt-function coefficients agree, providing another consistency check on the results presented in Section~\ref{subsec:3pt_results}.

\subsection{Towards holographic 4pt-functions}
\label{subsec:towards_4pt}

Let us now briefly turn to the computation of holographic 4pt-functions. The general procedure is to evaluate the gravitational on-shell action up to quartic order in the sources for boundary operators, and then take functional derivatives with respect to those sources. In practice, this amounts to evaluating all Witten diagrams allowed by the bulk supergravity couplings. For a generic 4pt-function of scalar operators, these are given by contact and exchange diagrams, where Lorentz symmetry restricts the exchanged field to be a scalar, vector or spin-2 fields. While conceptually straightforward, this approach becomes technically challenging when a large number of fields can be exchanged. One then needs to compute all relevant bulk couplings up to quartic order, evaluate the corresponding Witten diagrams and finally add up all contributions.

In our setup these calculations are somewhat simplified since we are dealing with a 5d supergravity consistent truncation and thus there cannot by any exchange Witten diagrams in which the exchanged field is a higher KK mode outside the consistent truncation. This greatly limits the number of diagrams, making the computation of 4pt-functions manageable. In the following, we showcase the simplifications which occur thanks to working in a consistent truncation by explicitly computing the correlators of the superconformal primary operators $\o_{\beta_1}$ and $\o_{\beta_2}$ in the UV $\mathcal{N}=4$ SYM theory, i.e. $\langle\o_{\beta_1}\o_{\beta_1}\o_{\beta_1}\o_{\beta_1}\rangle$, $\langle\o_{\beta_1}\o_{\beta_1}\o_{\beta_2}\o_{\beta_2}\rangle$ and $\langle\o_{\beta_2}\o_{\beta_2}\o_{\beta_2}\o_{\beta_2}\rangle$. The details of the computation are presented in Appendix~\ref{app:4pt_UV}, where we also verify that our results agree with the known 4pt-functions in the literature. Here we only summarise the two main advantages of working within the consistent truncation introduced above:
\begin{itemize}
\item First, there are indeed very few cubic and quartic terms involving the fields $\beta_1$ and~$\beta_2$. Expanding the effective Lagrangian \eqref{eq:Lagrangian_2}, the relevant couplings read
\begin{align}\label{eq:UV_bulk_couplings}
\begin{split}
	\Pt^{(3)} &= \frac{1}{2!}\Big(2\sqrt{\tfrac{2}{3}}\,\beta_1\beta_2^2\Big)+\frac{1}{3!}\Big(-2\sqrt{\tfrac{2}{3}}\,\beta_1^3\Big),\\[3pt]
	\Pt^{(4)} &\supset \frac{1}{2!\cdot2!}\Big(-\frac{2}{3}\beta_1^2\beta_2^2\Big) + \frac{1}{4!}\Big(-2(\beta_1^4+\beta_2^4)\Big).
\end{split}
\end{align}
and as noted before there are no derivative couplings for the scalars.
\item Second, it is clear from the structure of the Lagrangian \eqref{eq:Lagrangian_2} that all scalars couple only to expressions which are \textit{quadratic} in the gauge fields, which implies that there are no scalar-scalar-vector cubic couplings.\footnote{In other words, the scalars $\beta_{1,2}$ and $\phi_{1,2,3,4}$ are not charged under the ${\rm U}(1)$ gauge fields $\Am^{(J)}$ in this truncation.} Vector exchange diagrams are thus excluded in this model, and we are left with only scalar and graviton exchanges as well as contact diagrams, see Figure \ref{fig:4pt} for a graphical depiction.
\end{itemize}
%
\tikzset{snake it/.style={decorate, decoration=snake}}
\begin{figure}
\begin{center}
\begin{tikzpicture}[scale=0.5]
\begin{scope}
\draw (0, 0) circle (3);
\node[circle, fill=black, inner sep=1pt] (V) at (0,0) {};
\node[coordinate] (x4) at (2.12,2.12) {};
\node[coordinate] (x1) at (-2.12,2.12) {};
\node[coordinate] (x2) at (-2.12,-2.12) {};
\node[coordinate] (x3) at (2.12,-2.12) {};
\draw (V) -- (x1);
\draw (V) -- (x2);
\draw (V) -- (x3);
\draw (V) -- (x4);
\end{scope}
\begin{scope}[xshift=10cm]
\draw (0, 0) circle (3);
\node[circle, fill=black, inner sep=1pt] (V1) at (-1, 0) {};
\node[circle, fill=black, inner sep=1pt] (V2) at (1, 0) {};
\node[coordinate] (x4) at (2.12,2.12) {};
\node[coordinate] (x1) at (-2.12,2.12) {};
\node[coordinate] (x2) at (-2.12,-2.12) {};
\node[coordinate] (x3) at (2.12,-2.12) {};
\draw (V1) -- (x1);
\draw (V1) -- (x2);
\draw (V2) -- (x3);
\draw (V2) -- (x4);
\draw (V1) -- (V2);
\end{scope}
\begin{scope}[xshift=20cm]
\draw (0, 0) circle (3);
\node[circle, fill=black, inner sep=1pt] (V1) at (-1, 0) {};
\node[circle, fill=black, inner sep=1pt] (V2) at (1, 0) {};
\node[coordinate] (x4) at (2.12,2.12) {};
\node[coordinate] (x1) at (-2.12,2.12) {};
\node[coordinate] (x2) at (-2.12,-2.12) {};
\node[coordinate] (x3) at (2.12,-2.12) {};
\draw (V1) -- (x1);
\draw (V1) -- (x2);
\draw (V2) -- (x3);
\draw (V2) -- (x4);
\draw[decorate, decoration={snake, segment length=1.9mm, amplitude=1mm}] (V1) -- (V2);
\end{scope}
\end{tikzpicture}
\caption{Witten diagrams relevant for the computation of 4pt-correlators of $\o_{\beta_1}$ and $\o_{\beta_2}$. From left to right: contact diagram, $s$-channel scalar and graviton exchange diagrams.}
\label{fig:4pt}
\end{center}
\end{figure}
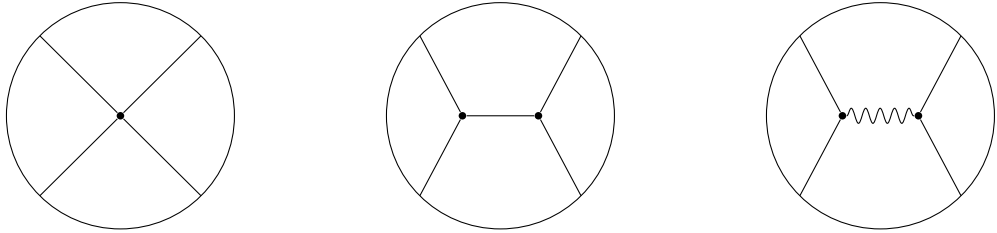
We believe that these two simplifications will also facilitate the computation of 4pt-functions in the IR theory. However, the evaluation of exchange diagrams in position space is technically more complicated when the conformal dimensions of the external or exchanged operators are non-integer, as is the case for some operators in the IR spectrum, c.f.~\eqref{eq:spectrum_scalar2}. We expect that working in Mellin space instead may offer the right language for the holographic computation of 4pt-functions in the LS SCFT. We leave this for future work \cite{BP_4pt}.

\section{Conclusions}
\label{sec:final}

In this work we used the holographic description of the LS 4d $\mathcal{N}=1$ SCFT to compute some of the 3pt-functions of BPS and non-BPS scalar operators in the planar limit of the theory. Our analysis serves as a blueprint for the calculation of holographic correlators in top-down AdS/CFT models with less than (half-)maximal supersymmetry. There are a number of open questions as well as possible generalizations and extensions of our work which we now briefly discuss.

\begin{itemize}
\item In our analysis we focused on the correlation functions of only 10 of the scalar operators in the LS SCFT dual to some of the lowest KK modes. Clearly, it is important to extend this analysis to all supergravity KK modes. It is straightforward, albeit very tedious, to perform this calculation for the remaining scalars in the 5d maximal gauged supergravity. The action of this theory is known explicitly and one needs to expand it to cubic order around the IR AdS$_5$ vacuum and then perform a very similar analysis to the one presented above. It is significantly more challenging to extend this calculation to the infinite tower of higher KK modes in type IIB supergravity. To this end one should probably adapt the techniques from Exceptional Field Theory developed in \cite{Duboeuf:2023cth} which are tailored to the calculation of the cubic supergravity couplings needed for the evaluation of the CFT correlation functions. It will be very interesting to perform this calculation explicitly and we hope that the results above, together with the subtleties about cubic derivative couplings and extremal correlators that we highlighted, will prove useful in this endeavor.

\item As we outlined in Section~\ref{subsec:towards_4pt}, the gauged supergravity truncations we studied can be used also to calculate 4pt-functions in the LS SCFT. It will be desirable to perform this calculation explicitly and also to extend it to 4pt-functions of the higher KK modes, perhaps by employing ExFT techniques or utilizing Mellin space methods. In addition to providing a wealth of new quantitative information about the planar limit of the LS theory, this analysis, combined with the methods reviewed in \cite{Bissi:2022mrs}, can also pave the way to studying $1/N$ corrections to local correlation functions.

\item The 3d ABJM SCFT at Chern-Simons level $k=1,2$ preserves $\mathcal{N}=8$ superconformal symmetry and admits a mass deformation that leads to an IR fixed point very analogous to the 4d LS SCFT that we studied here, see \cite{Benna:2008zy,Jafferis:2011zi,Bobev:2018uxk}. This mABJM IR SCFT preserves 3d $\mathcal{N}=2$ superconformal symmetry and has a well-known AdS$_4$ holographic dual that can be constructed in the 4d ${\rm SO}(8)$ maximal gauged supergravity \cite{Warner:1983vz} and subsequently uplifted to 11d supergravity. The spectrum of KK excitation around this AdS$_4$ vacuum solution can be calculated with ExFT methods, see \cite{Malek:2019eaz,Malek:2020yue}. As we will show in \cite{mABJM}, we can employ the 4d ${\rm SO}(8)$ gauged supergravity in a manner similar to the analysis in this work to compute some of the 3pt-functions in this 3d $\mathcal{N}=2$ holographic SCFT.

\item We are not aware of any QFT techniques that allow for the calculations of the 3pt-functions we analyzed holographically. Nevertheless, we hope that such methods may be developed in the future. Perhaps one can find supersymmetric localization calculations that will provide access to (integrated) correlation functions in 4d $\mathcal{N}=1$ SCFTs and this can then be used to compute 3pt-functions of some operators in the LS theory. One may also hope that some remnants of the integrable structure of the planar $\mathcal{N}=4$ SYM theory may remain intact after the RG flow to the LS SCFT and can be utilized to compute the correlators of interest. Importantly, such field theory calculations may provide a way to study the correlations functions of the LS theory across the whole conformal manifold and not only at the strongly coupled locus accessible by supergravity. It will be very exciting if any progress along these lines can be made.  

\item The RG flow between the $\mathcal{N}=4$ SYM theory and the LS SCFT is triggered by a relevant deformation that is universally present in any 4d $\mathcal{N}=2$ SCFT with an exactly marginal coupling \cite{Tachikawa:2009tt}. This universality is also reflected in holography where the RG flow is realized in many supergravity models, see \cite{Corrado:2002wx,Corrado:2004bz,Bobev:2018sgr}. This suggests that there may be some correlation functions in the UV and IR SCFTs that are captured by the holographic description and exhibit universal features. It will be very interesting to explore this question further and understand whether some of the 3pt-functions we calculated in this work enjoy such universal properties.

\item Our analysis has underscored the importance and convenience of utilizing 5d supergravity consistent truncations to calculate holographic correlators in top-down models of AdS/CFT. It should be possible to apply the same method to compute correlation functions in other AdS$_5$/CFT$_4$ examples. For instance, the 5d $\mathcal{N}=4$ consistent truncation of type IIB supergravity on $T^{1,1}$ derived in  \cite{Cassani:2010na} contains several scalar fields dual to operators in the Klebanov-Witten theory and thus can be readily used to compute their 3pt-functions. Recently, a 5d supergravity consistent truncation that captures a subset of the operators in 4d $\mathcal{N}=1$ and $\mathcal{N}=2$ class $\mathcal{S}$ SCFTs was derived in \cite{Bhattacharya:2024tjw}. Given our limited understanding of these intrinsically strongly coupled theories it will be most interesting to use the 5d supergravity model to compute their correlation functions.
\end{itemize}

\section*{Acknowledgments}
We acknowledge useful discussions with Lorenzo Bianchi, Shai Chester, Henriette Elvang, Fri\dh rik~Freyr~Gautason, Grisha Korchemsky, Marco Meineri, Krzysztof Pilch, Silviu Pufu, Shlomo Razamat, Slava Rychkov, Henning Samtleben, Inne Van de Plas, Jesse van Muiden, Xi Yin, Kostya Zarembo, and Xinan Zhou. We are grateful to Emanuel Malek and Alessandro Pini for the initial collaboration on this project and for the productive discussions. This work is supported in part by FWO projects G003523N, G094523N, and G0E2723N, as well as by the Odysseus grant G0F9516N from the FWO.  NB would like to thank the Institute for Advanced Study Princeton for the warm hospitality and inspiring atmosphere while part of this project was being completed.

\appendix
\section{3pt-functions in the UV }
\label{app:3pt_UV}
The purpose of this appendix is to reproduce -- within our 10-scalar model -- the classic result of \cite{Lee:1998bxa}, i.e. that the 3pt-functions of bulk scalars (dual to chiral primary operators) computed from supergravity on AdS$_5\times$S$^5$ precisely match those of \textit{free} $\N=4$ SYM theory in the large-$N$ limit. We proceed in two steps. First, we present the holographic computation following the general procedure outlined in Section~\ref{subsec:bulkcubic}. We then perform a purely field theoretic computation in the free $\N=4$ SYM theory, and show agreement with the results from holography.

\subsection{The holographic computation}
The starting point are the cubic couplings contained in the supergravity action expanded around the UV AdS$_5$ solution given in \eqref{eq:L_UV_exp}. Since there are no cubic terms coming from $\Lkin$, we have that all $d_{ijk}=0$. We therefore only need to consider the expansion of the potential, which at cubic order is given by
\begin{align}\label{eq:L_UV_cubic}
	\P_{\rm UV}^{(3)} = \frac{1}{2!}\Big(2\sqrt{\tfrac{2}{3}}\beta_1(\alpha_1^2+\alpha_2^2-2\alpha_3^2+\beta_2^2)+2\sqrt{2}\beta_2(\alpha_1^2-\alpha_2^2)\Big)+\frac{1}{3!}\Big(-2\sqrt{\tfrac{2}{3}}\beta_1^3\Big).
\end{align}
From here one reads off the bulk cubic couplings $c_{ijk}$. Using the results of Section \ref{subsec:bulkcubic}, we obtain the following set of non-vanishing 3pt-function coefficients:
\begin{align}\label{eq:3pt_UV}
\begin{split}
	C_{\alpha_1\alpha_1\beta_1}&=C_{\alpha_2\alpha_2\beta_1}=-C_{\beta_1\beta_1\beta_1}=C_{\beta_1\beta_2\beta_2}=\sqrt{\frac{2}{3}}\,\frac{1}{N}\,,\\
	C_{\alpha_1\alpha_1\beta_2}&=-C_{\alpha_2\alpha_2\beta_2}=\sqrt{2}\,\frac{1}{N}\,,\quad C_{\alpha_3\alpha_3\beta_1}=-2\sqrt{\frac{2}{3}}\,\frac{1}{N}\,,
\end{split}
\end{align}
where we used $\eta=\frac{N^2}{8\pi^2L^3}$ and we set $L=1$. These are the results of the bulk supergravity computation, which we now wish to reproduce from the dual field theory perspective.

\subsection{Comparison with free $\N=4$ SYM}

Recall from \eqref{eq:masses_UV} that in the UV the supergravity scalars $\alpha_{1,2,3}$ and $\beta_{1,2}$ are dual to scalar bilinear operators $\o_{\Phi_i}$, transforming in the $\mathbf{20'}$ of the $\rm SO(6)$ R-symmetry group. Their precise realisation in terms of the 6 adjoint scalars $X_I$ $(I=1,\ldots,6)$ of $\N=4$ SYM theory is given in \cite{Bobev:2016nua}, c.f. equations (4.28)-(4.29), and reads
\begin{align}\label{eq:operators1}
\begin{split}
	\o_{\alpha_i} &\sim \Tr(Z_iZ_i+\Zt_i\Zt_i)\,,\quad i=1,2,3,\\
	\o_{\beta_1}  &\sim \Tr(Z_1\Zt_1+Z_2\Zt_2-2Z_3\Zt_3)\,,\\
	\o_{\beta_2}  &\sim \Tr(Z_1\Zt_1-Z_2\Zt_2)\,,
\end{split}
\end{align}
where we have introduced the 3 complex fields $Z_i$ as
\begin{align}
	Z_1=\frac{X_1+{\rm i}X_2}{\sqrt{2}}\,,\quad Z_2=\frac{X_3+{\rm i}X_4}{\sqrt{2}}\,,\quad Z_3=\frac{X_5+{\rm i}X_6}{\sqrt{2}}\,,
\end{align}
and $\Zt_i$ denotes the complex conjugate. Next, we need to fix the normalisation of the above operators. In order to compute their norms, we use the Wick contractions
\begin{align}
	\langle (X_I)^{~b}_a(x_1)\, (X_J)^{~d}_c(x_2) \rangle = \delta_{IJ}\Big(\delta_a^d\delta_c^b-\tfrac{1}{N}\delta_a^b\delta_c^d\Big)g_{12}\,,\quad g_{12}\equiv\frac{1}{(x_1-x_2)^2}\,,
\end{align}
where $a,b,c,d$ are fundamental indices of the gauge group $\rm SU(N)$. In terms of the complex fields $Z_i$, we obtain
\begin{align}
	\langle (Z_i)^{~b}_a(x_1)\, (\Zt_j)^{~d}_c(x_2) \rangle = \delta_{ij}\Big(\delta_a^d\delta_c^b-\tfrac{1}{N}\delta_a^b\delta_c^d\Big)g_{12}\,,
\end{align}
and all other contractions vanish. Since the operators \eqref{eq:operators1} are bilinears involving terms of the form $\Tr(Z_iZ_i)$, $\Tr(\Zt_i\Zt_i)$, or $\Tr(Z_i\Zt_i)$, it is useful to first consider correlators of these simple building blocks. The non-vanishing 2pt-functions read
\begin{align}\label{eq:wick2}
\begin{split}
	\langle\Tr(Z_i\Zt_i)\Tr(Z_j\Zt_j)\rangle &= \delta_{ij}(N^2-1)g_{12}^2\,,\\
	\langle\Tr(Z_iZ_i)\Tr(\Zt_j\Zt_j)\rangle &= 2\delta_{ij}(N^2-1)g_{12}^2\,,
\end{split}
\end{align}
and for the 3pt-functions we have
\begin{align}\label{eq:wick3}
\begin{split}
	\langle\Tr(Z_i\Zt_i)\Tr(Z_j\Zt_j)\Tr(Z_k\Zt_k)\rangle &= 2\delta_{ij}\delta_{ik}(N^2-1)g_{12}g_{13}g_{23}\,,\\
	\langle\Tr(Z_i\Zt_i)\Tr(Z_j Z_j)\Tr(\Zt_k\Zt_k)\rangle &= 4\delta_{ij}\delta_{ik}(N^2-1)g_{12}g_{13}g_{23}\,.
\end{split}
\end{align}
Computing the 2- and 3pt-functions of the operators \eqref{eq:operators1} now reduces to taking linear combinations of \eqref{eq:wick2} and \eqref{eq:wick3}. The unit normalised operators then read
\begin{align}\label{eq:operators_normalised}
\begin{split}
	\o_{\alpha_i} &= \frac{1}{2\sqrt{N^2-1}}\Tr(Z_iZ_i+\Zt_i\Zt_i)\,,\quad i=1,2,3\,,\\
	\o_{\beta_1}  &= \frac{1}{\sqrt{6(N^2-1)}}\Tr(Z_1\Zt_1+Z_2\Zt_2-2Z_3\Zt_3)\,,\\
	\o_{\beta_2}  &= \frac{1}{\sqrt{2(N^2-1)}}\Tr(Z_1\Zt_1-Z_2\Zt_2)\,.
\end{split}
\end{align}
and one can check that they are orthogonal to each other. This is in agreement with the supergravity picture, where the kinetic terms are diagonal and hence different bulk scalars do not mix.

Moving on to the 3pt-functions, using \eqref{eq:wick3} we obtain
\begin{align}\label{eq:wick_3pt}
\begin{split}
	C^{(\N=4)}_{\alpha_1\alpha_1\beta_1}&=C^{(\N=4)}_{\alpha_2\alpha_2\beta_1}=-C^{(\N=4)}_{\beta_1\beta_1\beta_1}=C^{(\N=4)}_{\beta_1\beta_2\beta_2}=\sqrt{\frac{2}{3}}\,\frac{1}{\sqrt{N^2-1}}\,,\\
	C^{(\N=4)}_{\alpha_1\alpha_1\beta_2}&=-C^{(\N=4)}_{\alpha_2\alpha_2\beta_2}=\sqrt{2}\,\frac{1}{\sqrt{N^2-1}}\,,\quad C^{(\N=4)}_{\alpha_3\alpha_3\beta_1}=-2\sqrt{\frac{2}{3}}\,\frac{1}{\sqrt{N^2-1}}\,.
\end{split}
\end{align}
At leading order in the large-$N$ limit, these 3pt-function coefficients precisely match the holographic results given in \eqref{eq:3pt_UV} above.

\section{4pt-functions in the UV }
\label{app:4pt_UV}
In this appendix, we perform the holographic computation of 4pt-functions in the $\mathcal{N}=4$ SYM theory involving the operators $\o_{\beta_1}$ and  $\o_{\beta_1}$ by using the supergravity truncation introduced in Section~\ref{subsec:another_truncation}. We then compare the results with the known expressions in the literature, finding perfect agreement.

\subsection{The holographic computation}
The relevant bulk cubic and quartic couplings which involve the fields $\beta_1$ and $\beta_2$ are easily obtained by expanding the effective Lagrangian \eqref{eq:Lagrangian_2}. They are given by
\begin{align}\label{eq:UV_bulk_couplings2}
\begin{split}
	\Pt^{(3)} &= \frac{1}{2!}\Big(-2\sqrt{\tfrac{2}{3}}\,\beta_1\beta_2^2\Big)+\frac{1}{3!}\Big(2\sqrt{\tfrac{2}{3}}\,\beta_1^3\Big),\\[3pt]
	\Pt^{(4)} &\supset \frac{1}{2!\cdot2!}\Big(-\frac{2}{3}\beta_1^2\beta_2^2\Big) + \frac{1}{4!}\Big(-2(\beta_1^4+\beta_2^4)\Big).
\end{split}
\end{align}
With this at hand, one needs to evaluate the relevant 4pt tree-level Witten diagrams. The quartic couplings in~\eqref{eq:UV_bulk_couplings2} give rise to a contact diagram, $\A^{(c)}$. From the cubic terms, we see that only the scalars $\beta_{1,2}$ themselves can be exchanged, leading to scalar exchange diagrams which we denote by $\A^{(s)}$. Lastly, since all fields in the Lagrangian \eqref{eq:Lagrangian_2} couple to gravity, there is also a graviton exchange diagram, $\A^{(g)}$. Note that, as a special feature of the supergravity truncation we use, the scalars are not charged under the ${\rm U}(1)$ gauge fields and thus there is no contribution from vector exchange diagrams. For a graphical depiction of the relevant diagrams see Figure~\ref{fig:4pt}.

Recall that the fields $\beta_{1,2}$ are dual to dimension $\Delta=2$ operators, which have been extensively studied in the early days of the AdS/CFT correspondence. All of the above-mentioned diagrams are thus readily available in the literature, and are given in terms of so-called $\dbar{}$-functions, see e.g. \cite{Uruchurtu:2008kp}:
\begin{align}\label{eq:diagrams}
\begin{split}
	\A^{(c)} &= g_{12}^2g_{34}^2\,\cdot\,\frac{\pi^2}{2}\, u^2\,\dbar{2222}\,,\\
	\A_s^{(s)} &= g_{12}^2g_{34}^2\,\cdot\,\frac{\pi^2}{8}\, u\,\dbar{1122}\,,\\
	\A_s^{(g)} &= g_{12}^2g_{34}^2\,\cdot\,\frac{2\pi^2}{3}\, u\,(-2\dbar{1122}+3\dbar{2123}+3\dbar{2132}-3\dbar{3133})\,,
\end{split}
\end{align}
where $g_{ij}\equiv1/x_{ij}^2$ and $u$ is the usual conformal cross-ratio given by $u=\frac{x_{12}^2x_{34}^2}{x_{13}^2x_{24}^2}$. The subscript $s$ in the exchange diagrams denotes that these are the $s$-channel diagrams. The other two orientations, i.e. $t$- and $u$-channel diagrams, can be obtained by crossing transformations. The contact diagram does not have a preferred orientation since it is fully crossing symmetric.

Using the bulk couplings \eqref{eq:UV_bulk_couplings2}, we can now easily assemble the desired tree-level 4pt-functions. Beginning with the correlator of four scalars $\beta_1$, we have\footnote{The various Witten diagrams in \eqref{eq:diagrams} have been computed with unit normalised bulk-to-boundary propagators. Reinstating that normalisation factor (which for $\Delta=2$ operators is given by $\frac{1}{2\pi^2}$), and furthermore dividing by the square root of the 2pt-function to obtain the correlator of \textit{normalised} operators, yields the prefactor $1/(\sqrt{2\pi^2})^4$. Also, note that one has to include a relative minus sign between the exchange and contact diagrams since correlators are computed from $\Lkin-\P$ instead of $\Lkin+\P$.}
\begin{align}\label{eq:bulk_b1}
	\langle\o_{\beta_1}\o_{\beta_1}\o_{\beta_1}\o_{\beta_1}\rangle = \frac{1}{\eta(\sqrt{2\pi^2})^4}\bigg(\big[\A^{(g)}_s+\A^{(g)}_t+\A^{(g)}_u\big]+\frac{8}{3}\big[\A^{(s)}_s+\A^{(s)}_t+\A^{(s)}_u\big]+2\big[\A^{(c)}\big]\bigg),
\end{align}
where $\eta=\frac{N^2}{8\pi^2L^3}$. Next, for the correlator of four $\beta_2$'s the cubic and quartic terms in \eqref{eq:UV_bulk_couplings2} are such that we obtain exactly the same result, i.e.
\begin{align}\label{eq:bulk_b2}
	\langle\o_{\beta_2}\o_{\beta_2}\o_{\beta_2}\o_{\beta_2}\rangle = \langle\o_{\beta_1}\o_{\beta_1}\o_{\beta_1}\o_{\beta_1}\rangle
\end{align}
Lastly, the only non-vanishing mixed correlator is given by
\begin{align}\label{eq:bulk_b1b2mixed}
	\langle\o_{\beta_1}\o_{\beta_1}\o_{\beta_2}\o_{\beta_2}\rangle =\frac{1}{\eta(\sqrt{2\pi^2})^4}\bigg(\big[\A^{(g)}_s\big]-\frac{8}{3}\big[\A^{(s)}_s\big]+\frac{8}{3}\big[\A^{(s)}_t+\A^{(s)}_u\big]+\frac{2}{3}\big[\A^{(c)}\big]\bigg).
\end{align}

\subsection{Comparison with known results}

We now proceed with a simple consistency check on the above 4pt-functions by comparing them to known results in the literature. The supergravity computation of the 4pt-function of dimension $\Delta=2$ operators transforming in the $\mathbf{20'}$ irrep of ${\rm SO}(6)$ has been performed in \cite{Arutyunov:2000py}. Here we present their result in a more modern language. First, we introduce an index-free notation for the $\mathbf{20'}$ scalars by contracting the ${\rm SO}(6)$ R-symmetry indices with a null vector $y^I$:\footnote{Note that the operator $\o_2(x,y)$ as defined by \eqref{eq:o2_def} is not unit normalised and its 2pt-function reads
\begin{align}
	\langle\o_2(x_1,y_1)\o_2(x_2,y_2)\rangle = 2(N^2-1)\,\frac{(y_1\cdot y_2)^2}{x_{12}^4}\,.
\end{align}}
\begin{align}\label{eq:o2_def}
	\o_2(x,y) = y^Iy^J \Tr(X_IX_J)\,, \qquad y\cdot y\equiv y_I\,y^I = 0\,,
\end{align}
which automatically projects $\o_2(x,y)$ to transform in the symmetric traceless representation, i.e. the $\mathbf{20'}$ of ${\rm SO}(6)$. We can then write the 4pt-function as
\begin{align}\label{eq:o2_4pt}
	\langle\o_2(x_1,y_1)\o_2(x_2,y_2)\o_2(x_3,y_3)\o_2(x_4,y_4)\rangle = \frac{y_{12}^4y_{34}^4}{x_{12}^4 x_{34}^4}\times\mathcal{G}(u,v;\sigma,\tau)\,,
\end{align}
where $y_{ij}^2\equiv y_i\cdot y_j$ and we introduced the conformal and R-symmetry cross-ratios
\begin{align}\label{eq:cross-ratios}
\begin{split}
	u=\frac{x_{12}^2x_{34}^2}{x_{13}^2x_{24}^2}\,,\quad v=\frac{x_{14}^2x_{23}^2}{x_{13}^2x_{24}^2}\,,\quad \sigma=\frac{y_{13}^2y_{24}^2}{y_{12}^2y_{34}^2}\,,\quad \tau=\frac{y_{14}^2y_{23}^2}{y_{12}^2y_{34}^2}\,.
\end{split}
\end{align}
Second, using the partial non-renormalisation theorem of \cite{Eden:2000bk}, the supergravity result of \cite{Arutyunov:2000py} is simply given by
\begin{align}\label{eq:o2_4pt_sugra}
	\mathcal{G}_{\text{sugra}}(u,v;\sigma,\tau) = 16\,\Big(\sigma\, u+\tau\,\frac{u}{v}+\sigma\tau\,\frac{u^2}{v}\Big) + \mathcal{I}\times(-16u^2\dbar{2422})\,,
\end{align}
where $\mathcal{I}$ is fixed by superconformal symmetry and is given by the following polynomial of cross-ratios:
\begin{align}\label{eq:I_factor}
	\mathcal{I} = \tau + (1-\sigma-\tau)v+(\tau^2-\tau-\sigma\tau)u+(\sigma^2-\sigma-\sigma\tau)uv+\sigma v^2+\sigma\tau u^2\,.
\end{align}

In order to compare the result above with the corresponding calculations in the 10-scalar model, we need to write the operators $\o_{\beta_1}$ and $\o_{\beta_2}$ in terms of $\o_2(x,y)$ by choosing an appropriate set of polarisation vectors $y$. A simple way to solve the null constraint demanded by \eqref{eq:o2_def} is to consider vectors of the form $y^I=\{1,i,0,0,0,0\}$, and permutations of these entries. For convenience, let us define $y_{\{k,l\}}$ as the vector with entry $1$ at position $k$, and $i$ at position $l$, i.e. $y_{\{1,2\}}=\{1,i,0,0,0,0\}$ with complex conjugate vector $\yb_{\{1,2\}}=\{1,-i,0,0,0,0\}$. Using this notation one has
\begin{align}
	\o_2(x,y_{\{k,l\}})+\o_2(x,\yb_{\{k,l\}}) = 2\,\Tr(X_k^2-X_l^2)\,.
\end{align}
The operators of interest, $\o_{\beta_1}$ and $\o_{\beta_2}$, see \eqref{eq:operators_normalised} for their free-field realisation, can then be written as
\begin{align}\label{eq:beta1beta2_O2rep}
\begin{split}
	\o_{\beta_1} &= \frac{1}{4\sqrt{6(N^2-1)}}\sum_{y=y_{\{1,5\}},y_{\{2,5\}},y_{\{3,6\}},y_{\{4,6\}}}\left(\o_2(x,y)+\o_2(x,\yb)\right),\\
	\o_{\beta_2} &= \frac{1}{4\sqrt{2(N^2-1)}}\quad\quad\sum_{y=y_{\{1,3\}},y_{\{2,4\}}}\quad\quad\,\left(\o_2(x,y)+\o_2(x,\yb)\right).\\
\end{split}
\end{align}
Plugging these linear combinations into \eqref{eq:o2_4pt} and using the supergravity result \eqref{eq:o2_4pt_sugra}, one finds
\begin{align}\label{eq:4pt_results}
\begin{split}
	\langle\o_{\beta_1}\o_{\beta_1}\o_{\beta_1}\o_{\beta_1}\rangle &= \langle\o_{\beta_2}\o_{\beta_2}\o_{\beta_2}\o_{\beta_2}\rangle = \frac{1}{x_{12}^4x_{34}^4}\Big[\Big(u+\frac{u}{v}+\frac{u^2}{v}\Big) - (u+v+1)^2u^2\dbar{2422} \Big],\\[5pt]
	\langle\o_{\beta_1}\o_{\beta_1}\o_{\beta_2}\o_{\beta_2}\rangle &= \frac{1}{x_{12}^4x_{34}^4}\Big[\frac{1}{3}\Big(u+\frac{u}{v}+\frac{u^2}{v}\Big) - \frac{1}{3}(u^2-2uv-2u+v^2+10v+1)u^2\dbar{2422}\Big],
\end{split}
\end{align}
while all other mixed correlators vanish. Very nicely, this agrees with the previous results derived within the 10-scalar model: by using various identities between $\dbar{}$-functions, one can check that equations \eqref{eq:bulk_b1}-\eqref{eq:bulk_b1b2mixed} indeed simplify to \eqref{eq:4pt_results}.

\bibliographystyle{JHEP}
\bibliography{refs}
\end{document}